\documentclass[reprint, superscriptaddress, amsmath,amssymb,aps,prx]{revtex4-2}

\usepackage{physics}
\usepackage{graphicx}
\usepackage{dcolumn}
\usepackage{bm}
\usepackage{hyperref}
\usepackage{dsfont}
\usepackage{booktabs}
\usepackage{soul}
\usepackage{siunitx}
\usepackage{upgreek}
\hypersetup{
colorlinks=true,
linkcolor=cyan,
citecolor=magenta,
}

\newcommand{\nocontentsline}[3]{}
\newcommand{\tocless}[2]{\vspace{4ex}\bgroup\let\addcontentsline=\nocontentsline#1{#2}\egroup}

\begin{document}


\title{Bias-preserving and error-detectable entangling operations in a superconducting dual-rail system}

\author{Quantum Circuits, Inc. Team}\noaffiliation

\date{\today}

\begin{abstract}

To realize the potential of quantum computation, error-corrected machines are required~\cite{Gidneyfactoring2021} that can dramatically reduce the inevitable errors experienced by physical qubits. 
While significant progress has been made in reaching the surface-code threshold in superconducting platforms \cite{Googlebitflip2021,Googlesuppressing2023,Googlethreshold2024}, large gains in the logical error rate with increasing system size still remain out of reach. This is due both to the large number of required physical qubits and the need to operate far below threshold.
Importantly, by designing the error correction schemes to exploit the biases and structure of the physical errors, this threshold can be raised. 
Examples include codes that benefit from qubits with large Pauli-noise bias~\cite{puri_bias-preserving_2020,Mirrahimi_2014} and erasure qubits~\cite{Kubicadualrail2023,Teohdualrail2023,Kubicabiasednoise2023,Puribiasederasure2023,wu_erasure_2022}, in which certain errors can be detected at the hardware level.
Dual-rail qubits encoded in superconducting cavities are a promising erasure qubit wherein the dominant error, photon loss, can be detected and converted to an erasure. In these approaches, the complete set of operations, including two qubit gates, must be high performance and preserve as much of the desirable hierarchy or bias in the errors as possible. Here, we design and realize a novel two-qubit gate for dual-rail erasure qubits based on superconducting microwave cavities. The gate is high-speed ($\sim 500$~ns duration), and yields a residual gate infidelity after error detection below $0.1\%$, comparable to state-of-the-art results in any superconducting qubit platform.
Moreover, we experimentally demonstrate that this gate largely preserves the favorable error structure of idling dual-rail qubits, making it an excellent candidate gate for enabling error correction and fault-tolerant computing. We measure low erasure rates of approximately $0.5\%$\ per gate, as well as low and asymmetric dephasing errors that occur at least three times more frequently on control qubits compared to target qubits. Bit-flip errors are practically nonexistent and bounded at the few parts per million level. Such a high degree of error asymmetry has not been well explored but is extremely useful in quantum error correction and flag-qubit contexts, where it can create a faster path to error-corrected systems that rapidly suppress errors as they scale. 

\end{abstract}

\maketitle


\tocless\section{Introduction}
In order to progress from today's era of NISQ machines \cite{preskill_quantum_2018} and realize large error correction gains, the performance of physical qubits and their fundamental operations must improve significantly. Many physical platforms are now fairly mature, so that improvements in coherence times or in the control schemes tend to yield important but incremental advances. 
Another approach to relax the requirements for quantum error correction (QEC) is to utilize qubits with special noise properties and adapt the encoding scheme to match. However, to obtain the full benefit of this strategy, gate operations must also be designed to preserve as much of the error structure as possible \cite{aliferis_fault-tolerant_2008}.

For instance, one can engineer a qubit to achieve a large ratio between bit flips and phase flips, which can be realized by actively stabilizing a cat code  
\cite{PuriKerrcat2017,puri_bias-preserving_2020, grimm_stabilization_2020,putterman_hardware-efficient_2024, Mirrahimi_2014, Leghtas}
in a superconducting cavity. Such a biased-noise qubit could be employed with the XZZX or other rectangular variants of the surface code \cite{Tuckettbiasednoise2019,Tuckettbiasednoise2020,Puribiasednoise2021,Kubicabiasednoise2023,claes_tailored_2023}, which can in principle raise the threshold for gate errors from 1\% to more than 5\%. A five-qubit repetition code \cite{putterman_hardware-efficient_2024} using this approach was recently demonstrated to approach the break even point for an error-corrected memory. 

A second way to improve performance is to first detect errors as they occur at the hardware level.
For instance, if the dominant errors are leakage in the form of photon loss, as in photonic quantum computing \cite{KLM2001,PhotonicQIPreview2019}, the detectable loss of a qubit, called an erasure, can be very beneficial for error correction. For example, the well-known surface code has an increased threshold for phenomenological erasures of almost 25\% per round \cite{barrett_loss_errors_2010,delfosse_decoder_2020}, and can additionally correct twice as many erasures as Pauli errors \cite{grassl_codes_erasure_1997} for a given code distance. In neutral atoms and trapped ions, leakage to non-computational states can also be detected and converted to erasures \cite{wu_erasure_2022,ma_high-fidelity_2023,kangQuantumErrorCorrection2023}, with similar benefits \cite{Puribiasederasure2023,gu2024optimizingquantumerrorcorrection} if other sources of errors remain small. 

In superconducting circuits, dual-rail qubits using pairs of transmons \cite{ shim2016semiconductor,WillOliverDRTransmons,Kubicadualrail2023,Levinedualrail2024} or microwave cavities \cite{zakka2011quantum,Teohdualrail2023,chou_superconducting_2024} are paths to systems where the more benign erasures are the dominant error source. In the dual-rail cavity qubit, for instance, there is a favorable hierarchy of errors when idling. Erasures dominate by a factor of 5-10 over dephasing, and bit flips are several orders of magnitude rarer than dephasing events. So far, superconducting dual-rail cavity qubits have demonstrated fast, high fidelity single-qubit gates~\cite{chapman_beamsplitter_2023,lu_high-fidelity_2023}, good state preparation and measurement (SPAM)~\cite{chou_superconducting_2024}, and efficient, non-destructive detection of erasure errors~\cite{Akshaymidcircuit2024,degraafMidcircuitErasureCheck2025}. An important missing ingredient is an entangling operation between dual-rail cavity qubits, ideally one that is high performance and preserves the favorable error hierarchy. While there have been proposals~\cite{tsunoda_error_detectable_gates_2023}, to the best of our knowledge, such a gate has yet to be realized. 

Here, we propose and experimentally demonstrate a controlled-$Z$ (CZ) operation that preserves the error hierarchy in a system of superconducting dual-rail cavity qubits. This ``SWAP-Wait-SWAP" (SWS) gate uses high-fidelity parametric operations to temporarily swap an excitation from one of the cavities (that comprise the dual-rail control qubit) to a transmon coupler connecting the two qubits, similar to the gate implementation in ~\cite{RosenblumGao2018}. By temporarily populating the coupler, we are able to utilize the strong (MHz level) dispersive shift between the coupler and one cavity of the other (target) dual-rail qubit. Upon swapping the excitation back into the control cavity, we realize an entangling CZ gate in $\sim500$ nanoseconds.

Importantly, this excitation-preserving operation preserves the erasure-to-Pauli noise bias on both qubits, with erasure rates below 1\%. After erasure detection, we find the residual dephasing errors are at the level of 0.1\% per gate or below, as characterized by quantum state tomography and interleaved randomized benchmarking (RB). Moreover, we show that bit flips occur at less than a few parts per million per gate. 
Importantly, there is an additional error asymmetry wherein the target qubit erasures and dephasing rates are a factor of 3-4 times lower than those for the control qubit. This asymmetry can be beneficial when detecting error syndromes~\cite{teohlambdainprep}.
Finally, we study the error channel when one of the qubits suffers from photon loss during the gate, the dominant error in our system, and show both theoretically and experimentally that this amounts to a conditional-dephasing error on the unleaked qubit.
All of these properties can be exploited by properly designed error correction codes where the performance levels we demonstrate are well past the predicted surface code thresholds~\cite{teohlambdainprep, wu_erasure_2022}, and should thus allow significant logical error suppression with increasing surface code distance. 

\tocless\section{Results}
\begin{figure}[ht]
\centering
\includegraphics[width=3.3in]{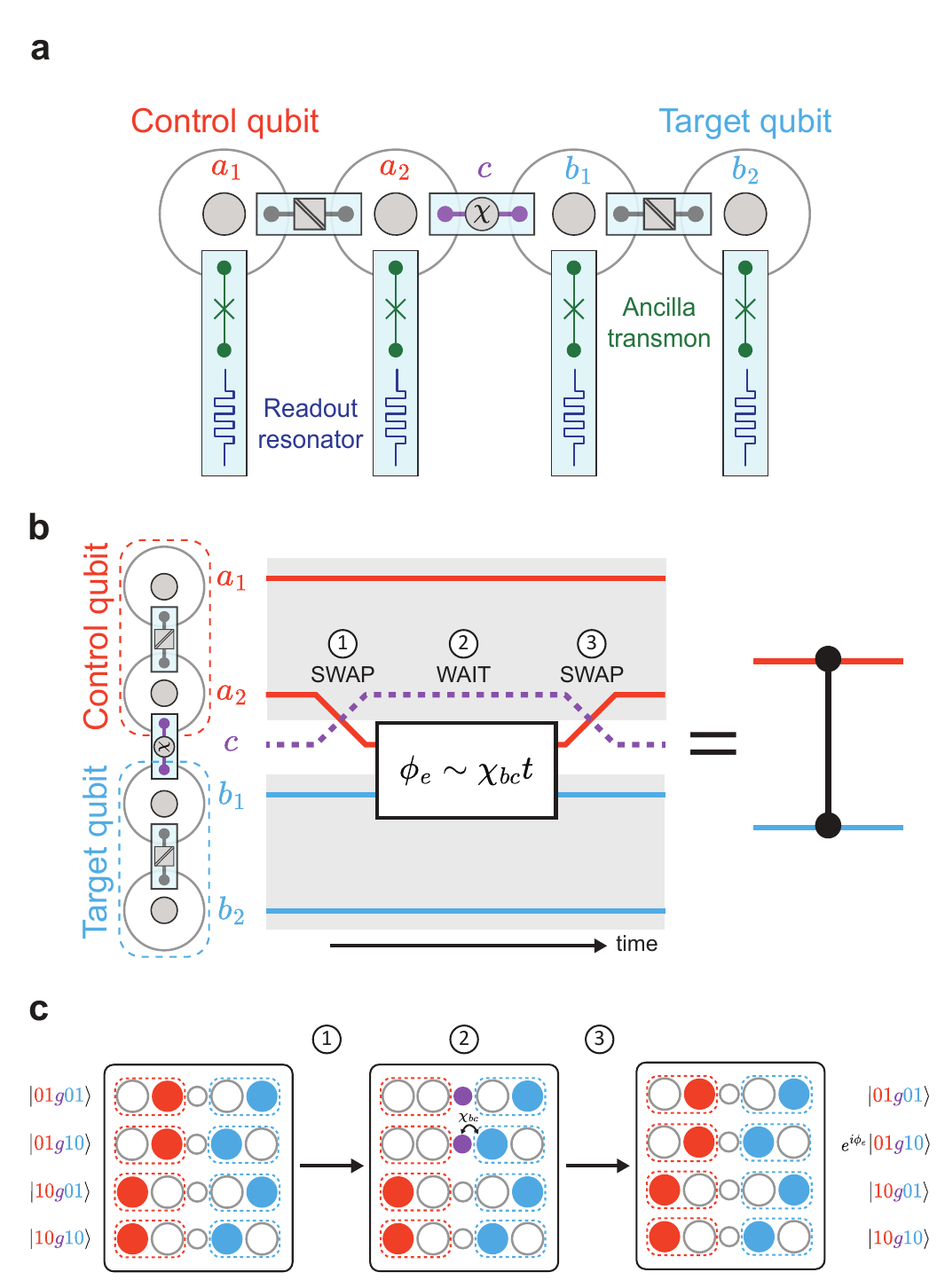}
\caption{
    \label{fig1}
    \textbf{Device hardware schematic and CZ gate sequence.}
    \textbf{a.} Schematic of a two dual-rail cavity qubit system, similar to ref.~\cite{chou_superconducting_2024}, where the control and target qubits are realized using pairs of three-dimensional $\lambda/4$ coaxial microwave cavities (red and blue). Three SQUID transmon couplers (grey and purple) each dispersively couple to pairs of cavity modes.  
    These couplers are flux-pumped~\cite{lu_high-fidelity_2023} to actuate parametric beamsplitter interactions for single qubit gates within the dual-rails, and also to implement the dual-rail two-qubit gate, which is achieved via the middle coupler. 
    Each cavity is coupled to an additional transmon qubit (green) and readout resonator (blue lines) that are used for state preparation and measurement.
    \textbf{b.} Diagram of the SWS gate sequence between two dual-rail cavity qubits. This gate consists of three steps: (1) switching on the dispersive interaction by performing a SWAP operation between the control qubit ($a_2$ cavity) and the coupler $c$, (2) allowing the dispersive interaction between coupler and target qubit ($b_1$ cavity) to acquire phase and entangle the two dual-rail qubits, and (3) switching off the dispersive interaction by performing a SWAP between control qubit ($a_2$ cavity) and coupler, restoring population to the $a_2$ cavity.
    \textbf{c.} Schematic illustration of photon population during the gate sequence for the four basis states.
}
\end{figure}
Our two-qubit gate makes use of a dispersive interaction to implement a CZ gate between two dual-rail qubits composed of nominally linear microwave cavities.
To explain our CZ gate, we first make a key observation: implementing an entangling interaction between the two nearest neighbor cavities results in entanglement between the two dual-rail qubits. Thus, only nearest-neighbor coupling is sufficient to implement the CZ gate. The physical system shown in Fig.~\ref{fig1}a realizes this paradigm by utilizing a coupler $c$ that interacts dispersively with both the $a_2$ and $b_1$ cavities. 
For our CZ gate construction we utilize the dispersive interaction between the coupler and the target qubit, $\mathcal{H}_{\mathrm{int}}/\hbar = \chi_{b c} \hat{b}_1^\dagger \hat{b}_1 \hat{c}^\dagger \hat{c}$, where $\chi_{bc}/2\pi=-1.51$ MHz in our system.

Normally, this static interaction term does not affect the system dynamics in any way, and is effectively `switched off' since the coupler mode $c$ is in its ground state at the beginning of the gate. 
The interaction can be effectively `switched on' by attempting to swap a photon between the cavity $a_2$ and coupler $c$, temporarily redefining the control dual-rail qubit between the non-local elements $a_1$ and $c$ (Fig.~\ref{fig1}b). 
Such a swap here is implemented using a parametric sideband interaction between the coupler and cavity: 
$\mathcal{H}/\hbar = \frac{g_{ac}}{2} \left(\hat{a}_2^\dagger \hat{c} + \hat{a}_2 \hat{c}^\dagger  \right)$, 
similar to the interaction used to effect single-qubit gates in dual-rails. In our hardware, $g_{ac}/2\pi = 4.23$ MHz. The dispersive interaction is then allowed to act for a time $t\sim\pi/\chi_{bc}$, thus entangling the two qubits.
A final swap between the coupler and cavity $a_{2}$ returns the excitation back to the control cavity. The entangling phase 
$\phi_e = \chi_{bc} t_{\mathrm{wait}}$ has thus been imparted only to the $\ket{01g10}$ initial state, resulting in a $\mathrm{CZ}$ gate~\footnote{Note that the phase of the two-qubit operation is parametrizable by adjusting the delay time, $t_{\mathrm{wait}}$. In this work, we always operate near $t\sim\pi/\chi_{bc}$ to enact $\mathrm{CZ}$ gates.}. We have introduced the notation $\ket{a_1,a_2,c,b_1,b_2}$ to describe the full state of the system in the number basis, including both dual-rail qubits and the coupler mode. 
Our gate construction allows us to effectively realize a large `switchable' dispersive interaction between cavity modes $a_2$ and $b_1$, without needing to modulate the strength of the underlying static dispersive interactions in our system Hamiltonian. Instead, we are making use of a simple SWAP operation to an auxiliary mode in order to actuate our entangling interaction, which is much easier than directly engineering a tunable `$\chi$' interaction in the system Hamiltonian~\cite{rosenblum2018fault}.

The construction of this two-qubit gate leads us to expect a simple error structure, which we observe and quantify below. This gate enables low error rates and gives rise to features that greatly benefit quantum error correction or flag-qubit error detection in quantum algorithms. 
First, the total number of excitations are preserved throughout the entire operation. That is to say, both control and target dual-rails never leave the single-excitation manifold (with a redefinition of the control dual rail during the gate operation to include the auxiliary coupler mode).  
This property ensures that any photon loss, even to the coupler mode,
leads to a detectable error for our CZ gate. 
Second, this gate scheme also avoids detrimental effects due to cavity self-Kerr, as no mode is populated with more than one photon at any point during the gate. 
Third, errors on the target and control qubits can have significant asymmetry. This is due the fact that only the control qubit is temporarily stored (if cavity $a_2$ is initially excited) in the lossy coupling transmon. The control qubit thus experiences 
increased relaxation (with probability $p=t / T_1\sim 1\%$) and dephasing (with $p=t / T_\phi\sim 0.1\%) $, where we have assumed $T_1\sim 50~\mu$s and $T_\phi\sim 500~\mu$s for the typical coherence times of a transmon. 
In contrast, the target dual-rail is simply left in its idling state throughout, so it has a smaller probability of undergoing a detectable leakage ($p\simeq t / T_{\rm 1, cav}\sim 0.1\%$) and dephasing ($p\simeq t /  T_{\rm \phi, DR}\sim 0.01\%$) given the idling single-photon decay times $T_{\rm 1, cav} \sim 500~\mu$s and dephasing times $T_{\rm \phi, DR} \sim 5$~ms. 
Fourth, we expect a strong suppression of bit-flip errors, as observed in the idling state of the dual-rails~\cite{chou_superconducting_2024}. This is because the outer cavities $a_1$ and $b_2$ and their couplers remain untouched during the gate. 
Finally, we expect well-behaved propagation of leakage errors. The dominant leakage channel is decay to the vacuum state, which switches off the CZ-generating cross-Kerr interaction. Thus, if either of the dual-rails relax to vacuum, subsequent CZ gates simply do not occur (the operation is effectively the identity).

This SWS gate can be calibrated via a straightforward procedure. In essence, it can be broken down into three pieces; i) swapping the photon from the control qubit cavity $a_2$ into the coupler $c$, ii) allowing the entangling phase from the dispersive interaction to accumulate over a specified delay time, and iii) swapping the photon back from the coupler into the control qubit cavity. In order to implement the CZ gate consisting of these pieces, we need to calibrate six parameters; the resonance frequency and duration of the parametric pump drive for cavity-coupler swap operation, the duration of the delay time needed to accumulate a $\pi$ entangling phase, the rotation angle for the second cavity-coupler swap to restore the photon population into the two-qubit manifolds, and the deterministic $Z$-phase of each single qubit gained during the gate. 
It is important to note that all these parameters can be independently calibrated in a sequential procedure; the one exception is the delay time, which requires prior adjustment of the rotation angle for the second cavity-coupler swap.
A more detailed explanation of these parameters and the calibration protocol can be found in Appendix~\ref{2Qcalib}.

\begin{figure*}[ht]
\centering
\includegraphics[width=\textwidth]{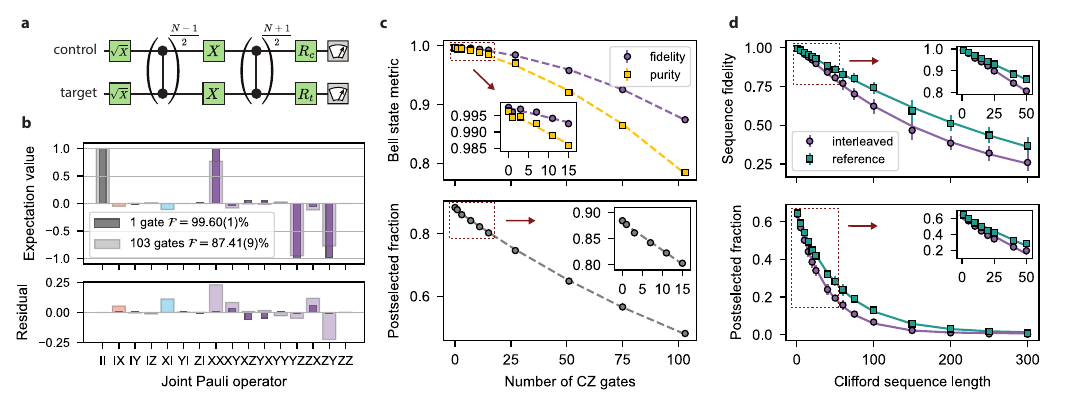}
\caption{
    \label{fig2}
    \textbf{Benchmarking a dual-rail qubit CZ gate.}
    \textbf{a.} Gate sequence used to quantify CZ gate fidelity via quantum state tomography. Applying an odd number $N$ of repeated CZ gates ideally generates a Bell state, and increases our sensitivity to gate errors. For all cases except $N=1$, $X$ gates are applied to each qubit midway through the sequence, after the first $(N-1)/2$ CZ gates. This serves to echo out low-frequency dephasing noise and to eliminate effects of no-jump backaction. By changing the single qubit gates, $R_c$ and $R_t$, we can extract all 16 joint-Pauli correlators for the final state.  
    \textbf{b.} Quantum state tomography after applying the gate sequence for $N=1$ CZ gates (dark) and $N=103$ CZ gates (light), represented by joint Pauli correlations. The ideal Bell state is defined by $\langle{II}\rangle=\langle{XX}\rangle=1$ and $\langle{YZ}\rangle=\langle{ZY}\rangle=-1$, with the residuals shown below.
    \textbf{c.} Repeated CZ gate Bell state tomography showing the increase in gate error with increasing gate repeats. We measure Bell state fidelity (upper panel, purple circles), state purity (upper panel, yellow squares), and postselected fraction, the fraction of shots for which no errors are detected (lower panel, grey), as a function of the number of CZ gates. Insets show data up to $N=15$ CZ gates. Error bars are generated from bootstrapped resampling of the data and are smaller than the size of the markers. 
    \textbf{d.} Two-qubit interleaved RB showing reference (blue squares) and interleaved (purple circles) experiments. Each Clifford sequence is randomized 20 times and the mean and standard deviation over random seeds are plotted. The fraction of shots measured in the two-qubit codespace are plotted in the lower panel.
}
\end{figure*}

Overall, our two-qubit system exhibits high fidelity SPAM and single-qubit gate fidelities advantageous for detailed characterization of our two-qubit gate. 
Single-qubit gates are performed by driving a beamsplitter interaction within the dual-rail for durations $\sim 200$~ns, achieving erasure rates of $\sim0.1\%$ and postselected RB error of $\sim0.01\%$. By using end-of-the-line erasure-detected measurements, we are able to obtain low postselected SPAM errors of $\sim0.02\%$~\cite{chou_superconducting_2024}.
We provide additional results in Appendix~\ref{app:1q_characterization}. 

We now present several methods for characterizing the fidelity and error structure of the SWS gate. As with all operations on erasure qubits, we must carefully quantify both the erasure rate and the residual Pauli errors. For our first demonstration of the SWS operation, we show high-fidelity entanglement between two dual-rail qubits. We implement the circuit shown in Fig.~\ref{fig2}a with $N=1$, which generates a Bell state. We subsequently perform quantum state tomography using an overcomplete gateset $\{I, X_{\pm \frac{\pi}{2}}, X, Y_{\pm \frac{\pi}{2}}\}$. As shown in Fig.~\ref{fig2}, we generate a high-fidelity Bell state and, after postselection of erasures, extract an end-to-end circuit state fidelity of $99.60(1)\%$ and a state purity of $99.46(3)\%$. 
We highlight that this infidelity of $\sim 0.4\%$ includes errors aggregated from SPAM, single-qubit gates and the CZ gate.

Next, we perform repeated CZ gates to establish a more precise metric for the errors-per-CZ-gate, using the sequence shown in Fig.~\ref{fig2}a. We apply an odd number ($N$) of total CZ gates to repeatedly entangle and disentangle the two qubits and subsequently characterize the resulting Bell state with quantum state tomography. We also include an $X$ gate after the first $(N-1)/2$ CZ gates which serves to eliminate the effect of no-jump backaction, first-order drift in control signal amplitudes, and any extraneous sources of low-frequency dephasing noise. We perform up to $N=103$ repeated gates and show the corresponding state tomography results in Fig.~\ref{fig2}b, extracting an end-to-end postselected state fidelity of $87.4(1)\%$ and a state purity of $78.4(2)\%$. From this we infer, after postselecting erasures and with no SPAM corrections, a bound of 0.12\% on the remaining error-per-gate, among the best reported for superconducting qubits. 
The dependence on number of gates is shown in Fig.~\ref{fig2}c, where we present state fidelity, state purity, and postselected fraction as a function of the number of repeated CZ gates. 
First, we can consider the behavior at lower gate repeats. As shown in the inset of Fig.~\ref{fig2}c, our data shows a linear decrease in fidelity and purity out to  $N=15$ repeated gates, from which we extract a postselected infidelity of $0.03\%$ per gate. 
For larger gate depths, both the fidelity and purity of the final state display a quadratic decrease, which may be due to drift in calibration parameters, or fluctuations in the frequency of the coupling transmon on timescales of experimental shot averaging and will require further investigation. Adding more than one $X$ gate does not help to alleviate this effect. Finally, from our tomography data we also extract the postselected fraction, defined to be the fraction of experimental shots that were not assigned to an erasure. From fitting the results to an exponential decay, we extract a total erasure probability per CZ gate of $0.53(2)\%$.

As an alternative method for determining the postselected gate fidelity, we perform interleaved RB to test our CZ gate (Fig.~\ref{fig2}d). RB in the presence of leakage errors sometimes requires taking into account the nuanced effects \cite{RBleakage, wood_rb_leakage_2018} of leakage and seepage, where leaked states can silently re-enter the codespace. However, given our hardware error rates these effects are negligible.  We use the gateset $\{X_{\frac{\pi}{2}}, Z_{\pm\frac{\pi}{2}}, Z, \mathrm{CZ}\}$ to generate the full two-qubit Clifford group, performing 20 randomized sequences for each Clifford sequence length. Because of the finite erasure probability ($0.69\%$) per gate, the usable sequence depth is limited to only a few hundred gates, as shown in the postselected fraction curve in the lower panel. This complicates the fitting to the usual exponential form expected for the sequence yield (top panel of Fig.~\ref{fig2}d). We therefore use a linear fit to the curves for shorter depths (inset), obtaining an estimate for the error per CZ gate of $0.1\%$, in reasonably good agreement with the estimate from repeated Bell state tomography.

\begin{table*}\label{tab:error_hierarchy}
    \centering
\begin{tabular}{| c | c | p{2cm}| p{2cm} | p{6cm} |}
    \hline \hline
    Error hierarchy & Total cost per CZ gate & \multicolumn{2}{c|}{Error asymmetry (control:target)} & Error propagation for QEC \\
                    &                       & \multicolumn{1}{c|}{Typical} & \multicolumn{1}{c|}{Measured} & \\
    \hline
    Erasure          & $\pi / \left(\chi_{bc} T_1^{\mathrm{c}}\right)  \sim 0.5\%$ & \centering 10:1 & \centering 4:1 & $Z$-error on unleaked qubit \\
    Dephasing        & $\pi / \left( \chi_{bc} T_{\phi}^{\mathrm{c}}\right) \sim 0.1\%$ & \centering 10:1 & \centering 3.5:1 & Commutes with gate; ancilla dephasing becomes stabilizer measurement error \\
    Bit-flip         & $\lesssim 10^{-6}$ & \centering 10:1 & \centering  immeasurable & Very rare, contributes to circuit-level depolarizing noise  \\
    \hline \hline
\end{tabular}
    \caption{Error hierarchy and asymmetry for the SWS gate. Values reflect both the typical and measured  performance expected for a SWS gate and their effect on quantum error correction schemes.} 
    \label{tab:error_table}
\end{table*}

\begin{figure}[ht]
\centering
\includegraphics[width=\columnwidth]{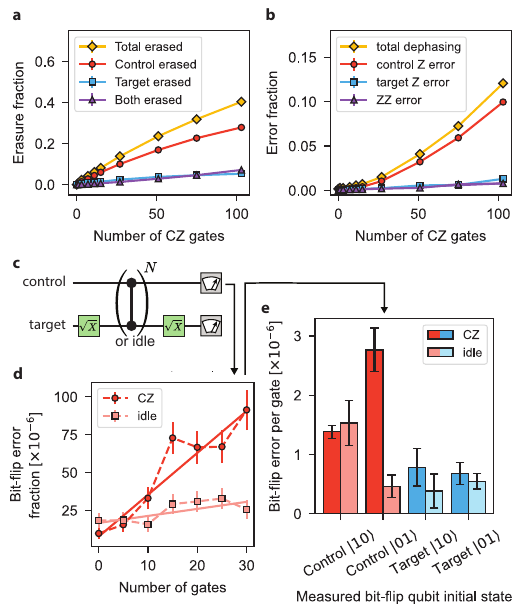}
\caption{
    \label{fig3}
    \textbf{Quantifying CZ error structure.}
    \textbf{a.} Extracted erasure fraction and \textbf{b.} $Z$-error fraction after applying a variable number of CZ gates, showing a breakdown for the following categories of dephasing errors: control only (red circles), target only (blue squares), both control and target (purple triangles), and total error (yellow diamonds). For both results, a strong asymmetry between the control and target erasures is observed.
    \textbf{c.} Example circuit to measure control qubit bit-flip errors during $N$ CZ gates or the equivalent idle duration. A pair of experiments are performed with the control qubit initialized in either $\ket{01}$ or $\ket{10}$. The circuit for measuring target qubit bit-flip errors is similar with $\sqrt{X}$ gates performed instead on the control qubit.
    \textbf{d.} Measured control qubit bit-flip fraction when initialized in $\ket{01}$ as a function of number of applied CZ gates (circles) and as a function of the equivalent idle (squares). From a linear fit to the data, we infer a total bit-flip error per CZ gate of $2.8(4)\times 10^{-6}$ and $0.5(2)\times 10^{-6}$ when simply idling.
    \textbf{e.} Extracted bit-flip error fractions per CZ gate for control qubit (red, left) and target qubit (blue, right). CZ bit-flip errors are compared against idling errors, showing low bit-flip errors at the level of $\sim10^{-6}$ per gate.
}
\end{figure}

Having benchmarked the performance of the CZ gate, we now investigate the structure of gate errors on the control and target qubit.
Because our CZ gate relies on swapping of the control qubit into the coupler and back, for part of the CZ gate, the control qubit will be physically defined between one cavity and the coupler. As such, gate errors, such as leakage and dephasing, will be largely set by the coupler coherences. On the other hand, the target qubit simply idles for the duration of the gate and so leakage and dephasing will be set by its longer idling coherences. With our system parameters we observe about 3.5 times lower errors on the target qubit than on the control qubit. Counterintuitively, this error asymmetry can be a highly beneficial property, particularly in an error-correction context. For example, when performing error-correction one may assign the ancilla (data) qubit to always correspond to the control (target) qubit. With this choice, data qubits will suffer lower dephasing errors, and ancilla qubits errors will just lead to syndrome measurement errors. Furthermore, bit-flip errors are practically nonexistent in our CZ gate, which greatly limits the injection of errors into the data qubits of the logical qubit. In the following, we measure the error asymmetry for both leakage and dephasing errors as well as bit-flip errors. 

First we characterize the fraction of detected leakages for our repeated gate Bell state tomography, extracting assigned erasures for the control qubit only, target qubit only, and both qubits, shown in Fig.~\ref{fig3}b. As expected we observe a higher fraction of erasures on the control qubit compared to the target qubit, extracting $0.400(4)\%$ and $0.096(4)\%$ per CZ gate for control and target qubit, respectively. Taken together, this erasure asymmetry of $\sim 4$ is consistent with the relative $T_1$ between the coupler and the average $T_1$ of the target qubit cavities.

Next, we study the relative fraction of dephasing errors on the control and target qubit. Analyzing the reconstructed quantum state from our Bell state tomography, we extract $Z$ errors on the target and control qubit individually as well as correlated $ZZ$ dephasing on both qubits.
Our results are shown in Fig.~\ref{fig3}c, showing a clear asymmetry in $Z$-errors between the control and target qubit, and smaller $ZZ$ errors. From these results, we extract target qubit $Z$ errors of $0.0112(9)\%$ per CZ gate. The control qubit $Z$ errors show the same type of nonlinear behavior observed for the QST results, suggesting that the root cause of this behavior could be attributed to some characteristic of the coupler. To quantify the control qubit error, we evaluate the error at large $N$, choosing $N = 27$ repeats and extracting $Z$-errors of $0.039(1)\%$ per gate. Taken together, we observe a dephasing ratio of $\sim 3.5$. These results are consistent with a simple model where dephasing errors on the control qubit are set by coupler $T_\phi$ as we swap into the coupler during the CZ gate while target qubit dephasing is set by the dual-rail $T_\phi$, and this analysis confirms this novel error structure for our CZ gate.

Finally, we measure the fraction of bit-flip errors while performing the CZ gate. From \cite{chou_superconducting_2024}, we expect idling bit-flip errors to be small and here we measure the added bit-flip error from executing the CZ gate. We perform the following measurement protocol for measuring bit-flip errors on the control (target) qubit. We initialize the control (target) in a basis state, $\ket{01}$ or $\ket{10}$, and perform a Ramsey-style circuit on the target (control) qubit with $N$ CZ gates sandwiched between the $\pi/2$ pulses. We measure the fraction of bit-flip outcomes on the control (target) qubit. These results are summarized in Fig.~\ref{fig3}d-f. As apparent from the figure, the overall probability of bit flips is bounded at few parts per million per gate, probably limited by leakage misassignments. 
We note that although the CZ gate preserves the Pauli noise-bias, we do not anticipate using a full gate set that preserves Pauli noise-bias. 

\begin{figure*}[ht]
\centering
\includegraphics[width=\textwidth]{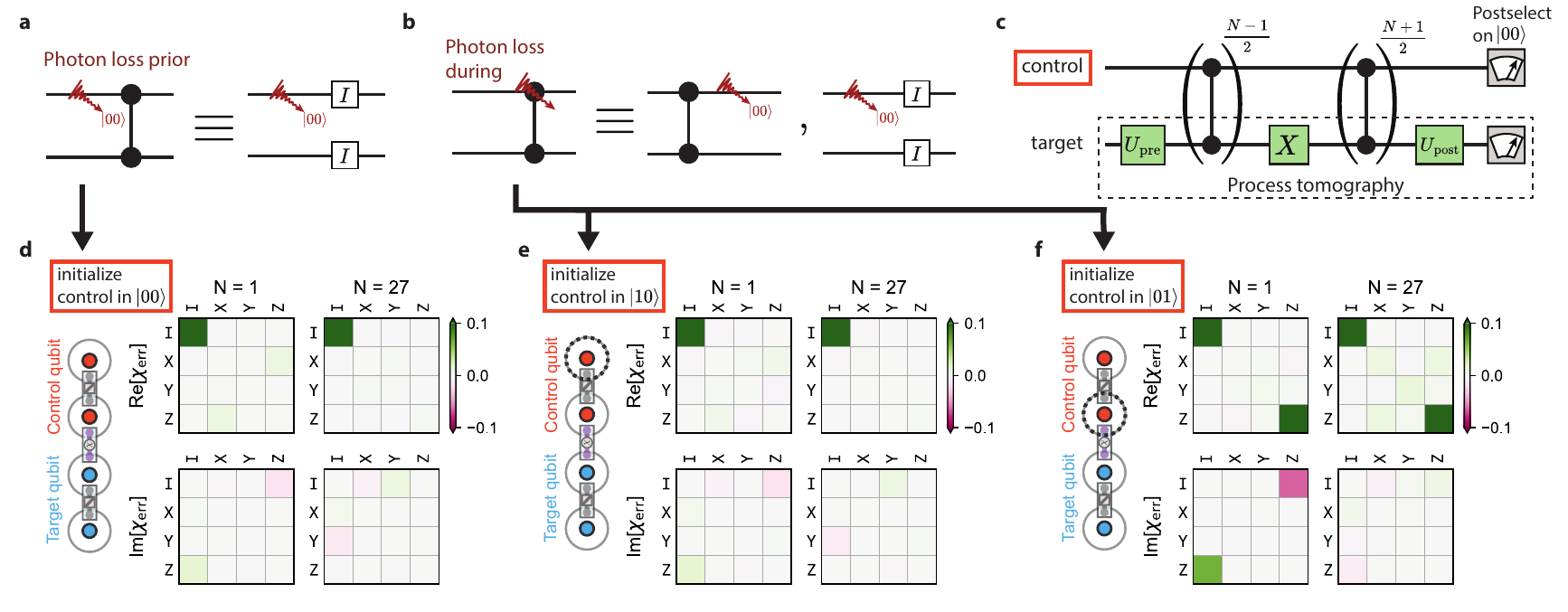}
\caption{
    \textbf{Leakage propagation properties of the CZ gate.}
    \textbf{a.} Leakage on the control qubit to $\ket{00}$ prior to the CZ gate causes the gate to be switched off and instead the identity process is performed on both qubits.
    \textbf{b.} If photon loss occurs during the gate, the error channel can be modeled as either leakage occurring before (and the gate is switched off as above) or after the CZ gate (resulting in a CZ error). 
    \textbf{c.} Circuit diagram used to test the leakage propagation properties of the CZ gate. For different preparations on the control qubit, quantum process tomography is performed on the target qubit after applying $N$ CZ gates on the target qubit. To test for control qubit leakage properties, the tomography is conditioned on the control qubit being found in the leakage $\ket{00}$ state. 
    \textbf{d.-f.} Conditioned process tomography on target qubit when the control qubit is prepared in $\ket{00}$, $\ket{10}$, and $\ket{01}$, respectively. For each case, the real (upper) and imaginary (lower) part of the $\chi$-error matrices are plotted for $N=1$ CZ gates (left) and $N=27$ CZ gates (right). The color scale is clipped at 0.1 to highlight deviations from the expected error channel. For control qubit initialized in $\ket{00}$ and $\ket{10}$, the expected identity process is observed indicating that a leakage error turns off the gate; while the expected `CZ error' is observed for leakage when the control qubit is initialized in $\ket{01}$ indicated by large diagonal components for $I$ and $Z$ in the real part of $\chi$-error matrix.}
\label{fig4}
\end{figure*}

Leakage is typically considered to be a particularly damaging error in a quantum error correction context. If not properly controlled, leakage may propagate without bounds until error correction fails. For our dual-rail qubits, leakage will be quickly detected and converted to erasures via our erasure check measurements. However, we do not intend to use erasure checks after every gate operation and so we investigate how the dominant $\ket{00}$ leakage state propagates through our CZ gates while awaiting detection. 

In our system, we consider two possible leakage scenarios: photon loss to $\ket{00}$ that occurs prior to a CZ gate and leakage that occurs mid-way through the CZ gate at an unknown time. Furthermore, we study the more likely scenario of leakage affecting only the control qubit, although due to the symmetry of 
a CZ gate our results are expected to apply similarly to the case of leakage on the target qubit. Whilst other dual-rail leakage states are possible due to cavity heating, such as $\ket{02}$ or $\ket{11}$, these occur at least $1000$ times less frequently in our hardware and are not considered here. 

To see why leakage propagation is particularly benign, we can first consider what happens if the control qubit is in the $\ket{00}$ state prior to the CZ gate. Since there is only one excitation in the system (in the target qubit), the dispersive interaction is never activated resulting in an identity operation on the target qubit. The exact same scenario occurs if the control qubit begins in $\ket{10}$ and suffers from photon loss during the CZ gate: since no photons were ever swapped into the coupler, the dispersive interaction was never activated which again results in an identity operation on the target qubit. The interesting case occurs when the control qubit begins in $\ket{01}$ and swaps into the coupler. Should excitation loss occur at any time during the gate, the result will be an incomplete CPHASE gate with unknown rotation angle. Averaging over all possible loss times results in what we term a `CZ error', a correlated error in which the target qubit dephases, but only if the control leaks from the initial $\ket{01}$ state. Such an error model is useful for enabling delayed erasure checks. A more detailed and theoretical treatment of the leakage error model can be found in Appendix \ref{app:leakage_prop}.

The following experiments characterize and verify the aforementioned CZ gate leakage propagation properties, shown in Fig.~\ref{fig4}c.
We perform $N$ repeated CZ gates followed by process tomography on the target qubit, for several different state preparations on the control qubit. Importantly, we postselect on measuring the control qubit in the leakage state $\ket{00}$ in all cases. To quantify the error channel, we plot the leakage-conditioned quantum process tomography using the the $\chi$-error matrix~\cite{korotkov_error_2013} against a reference process, here the identity channel. If the measured process is free of errors, then the error matrix will equal the identity process, with just a single non-zero element. From this representation it is easy to extract the Pauli error channel on the (unleaked) target qubit, which is given by
\begin{equation}
    \Lambda(\rho) = \sum_{ij}\chi_{ij}^{\mathrm{err}} \hat{\sigma}_i \rho \hat{\sigma}_j,
\end{equation}

We first consider the two leakage cases when initializing the control qubit in the $\ket{00}$ and $\ket{10}$ state. Our results, shown in Fig.~\ref{fig4}d,e for $N=1$ and $N=27$ CZ gates, show the expected behavior, that the resulting process on the target qubit is close to the identity process. We use the $\chi$-error matrix and quantify the deviation from the identity process by extracting the residual total Pauli error. For $N=1$, we extract a small residual Pauli error of $0.19(2)\%$ and $0.18(2)\%$ for $\ket{00}$ and $\ket{10}$ preparation, respectively. By repeating the gate many times, we are able to increase the fraction of erasures that occurred during the CZ gates over those that occurred during SPAM. At $N=27$ we estimate the fraction of CZ erasures to be roughly $2/3$ of total erasures. We extract a still small residual Pauli error of $0.33(2)\%$ and $0.35(2)\%$, experimentally confirming that these types of leakage errors essentially turn off the gate. 

Finally, we consider the case when the control qubit is in $\ket{01}$ and the CZ interaction is turned on for a random duration until the leakage event turns off the interaction. The resulting process on the target qubit is shown in Fig.~\ref{fig4}f, and closely matches a pure dephasing channel. For $N=1$ gates the extracted residual $X$ and $Y$ errors total $0.7(1)\%$ and at $N=27$ this error totals $2.55(5)\%$, a number that quantifies the deviation of this error channel from the theoretically expected dephasing channel. Dividing this number by $N$ yields $0.094(2)\%$, a number that characterizes an upper bound for the $X$ and $Y$ Pauli errors experienced by the target qubit when the control qubit leaks during a CZ gate. 

\tocless\section{Discussion and conclusion}
In summary, we have devised and demonstrated what is to our knowledge the first entangling gate for superconducting erasure qubits. This SWS gate uses the existing transmon parametric couplers to allow sub-microsecond gate times, and achieves state-of-the-art performance. Moreover, it preserves (see Table \ref{tab:error_table}) the desirable error hierarchy and long coherence times of the dual-rail cavity qubit. By virtue of being an excitation preserving operation, the dominant errors ($\sim 0.5\%$) remain erasures which can be detected by end-of-the-line \cite{chou_superconducting_2024} or mid-circuit \cite{Akshaymidcircuit2024,degraafMidcircuitErasureCheck2025} erasure check measurements. The high postselected fidelity (99.97-99.90\%) is then primarily limited by dephasing in the dual-rail and the transmon coupler, the strong bias against bit flips is maintained, which are bounded by our measurements to be less than a few parts per million per gate. The gate displays a significant asymmetry in the errors induced on the control (i.e. the dual-rail that swaps into the coupler) and the target qubits. Remarkably, we observe that the Pauli error-per-gate on the target qubit (after postselection) is as low as 0.01\%. Finally, the gate has a very simple behavior when one of the qubits is already erased or suffers an erasure during the gate, which should minimize the propagation of errors and reduces the frequency of mid-circuit erasure checks. Further improvements in performance by increasing the gate speed or the coherences of the coupler and dual-rails may be possible in future implementations.  

The addition of a fast and high-fidelity entangling operation completes the gate set for universal operations for the dual-rail cavity qubit, and can enable high performance in short-depth circuits and new types of error mitigation. In addition, the properties of SWS gate are highly optimal in the context of quantum error correction. For example, the erasure and dephasing rates are a factor of ten below their respective separate thresholds for the surface code. In the presence of both kinds of errors, detailed numerical simulations~\cite{teohlambdainprep} are needed to estimate the actual performance, but the results described here provide a realistic way to realize large error correction gains of 10 or more with increase in code distance in the near future, making the superconducting dual-rail cavity qubit an attractive path for achieving fault-tolerant quantum computing.

\tocless\section{Acknowledgments}
We would like to thank Luigi Frunzio, Sophia Xue, John Garmon, and Takahiro Tsunoda for their feedback on the manuscript. We would also like to acknowledge Ray Smets for his executive leadership and support for the entire Quantum Circuits team.

\newpage
\tocless\section{Author list} 
{Quantum Circuits, Inc. Team and Collaborators}

\bigskip
{
\renewcommand{\author}[2]{#1$^\textrm{\scriptsize #2}$}
\renewcommand{\affiliation}[2]{$^\textrm{\scriptsize #1}$ #2 \\}

\newcommand{\xQCI}{\affiliation{1}{Quantum Circuits, Inc., New Haven, CT, USA}}
\newcommand{\QCI}{1}

\newcommand{\xYalePhysics}{\affiliation{2}{Departments of Applied Physics and Physics, Yale University, New Haven, CT, USA
}}
\newcommand{\YalePhysics}{2}

\newcommand{\xYaleQuantumInstitute}{\affiliation{3}{Yale Quantum Institute, Yale University, New Haven, Connecticut, USA
}}
\newcommand{\YQI}{3}

\newcommand{\xMicrosoft}{\affiliation{4}{Current address: Microsoft Quantum
}}
\newcommand{\Microsoft}{4}

\newcommand{\xWeizmann}{\affiliation{5}{Current address: Department of Condensed Matter Physics, Weizmann Institute of Science, Rehovot, Israel
}}
\newcommand{\Weizmann}{5}

\newcommand{\xEqualContribution}{\affiliation{*}{Equal Contribution}}
\newcommand{\EqualContribution}{*}

\author{Nitish Mehta}{\QCI,\EqualContribution},
\author{James D. Teoh}{\QCI,\EqualContribution},
\author{Taewan Noh}{\QCI,\EqualContribution},
\author{Ankur Agrawal}{\QCI},
\author{Amos Anderson}{\QCI},
\author{Beau Birdsall}{\QCI},
\author{Avadh Brahmbhatt}{\QCI},
\author{Winfred Byrd}{\QCI},
\author{Anthony Cabrera}{\QCI},
\author{Marc Cacioppo}{\QCI},
\author{Leo Carroll}{\QCI},
\author{Jonathan Chen}{\QCI},
\author{Tzu-Chiao Chien}{\QCI,\Microsoft},
\author{Richard Chamberlain}{\QCI},
\author{Jacob C. Curtis}{\QCI},
\author{Doreen Danso}{\QCI},
\author{Sanjana Renganatha Desigan}{\QCI},
\author{Francesco D’Acounto}{\QCI},
\author{Bassel Heiba Elfeky}{\QCI},
\author{S. M. Farzaneh}{\QCI},
\author{Chase Foley}{\QCI},
\author{Benjamin Gudlewski}{\QCI},
\author{Hannah Hastings}{\QCI},
\author{Robert Johnson}{\QCI},
\author{Nishaad Khedkar}{\QCI,\YalePhysics,\YQI},
\author{Trevor Keen}{\QCI},
\author{Anup Kumar}{\QCI},
\author{Cihan Kurter}{\QCI},
\author{Kamila Krawczuk}{\QCI},
\author{Eric Langstengel}{\QCI},
\author{Richard D. Li}{\QCI},
\author{Gangqiang Liu}{\QCI},
\author{Hanyi Lu}{\QCI},
\author{Pinlei Lu}{\QCI},
\author{Luke Mastalli-Kelly}{\QCI},
\author{Adam Maines}{\QCI},
\author{Michael Maxwell}{\QCI},
\author{Heather McCarrick}{\QCI},
\author{Mona Mirzaei}{\QCI},
\author{Anirudh Narla}{\QCI,\Microsoft},
\author{Omar Rashad}{\QCI},
\author{Erik Reikes}{\QCI},
\author{Mizanur Rahman}{\QCI},
\author{Rurik Primiani}{\QCI},
\author{Michael Schwaller}{\QCI},
\author{Ali Sabbah}{\QCI},
\author{Tali Shemma}{\QCI,\Weizmann},
\author{Ruby A. Shi}{\QCI},
\author{Sitakanta Satapathy}{\QCI},
\author{Dean Stolpe}{\QCI},
\author{Jonathan Strenczewilk}{\QCI},
\author{Doug Szperka}{\QCI},
\author{Iu-Wei Sze}{\QCI},
\author{David Sweeney}{\QCI},
\author{Preetham Tikkireddi}{\QCI},
\author{Chin-Lun Tsung}{\QCI},
\author{Daren Vet Sam}{\QCI},
\author{Daniel K. Weiss}{\QCI},
\author{Zhibo Yang}{\QCI},
\author{Liuqi Yu}{\QCI},
\author{Teng Zhang}{\QCI},
\author{Olivier Boireau}{\QCI},
\author{Stephen Horton}{\QCI},
\author{Sean Weinberg}{\QCI},
\author{Jos\'e Aumentado}{\QCI},
\author{Bryan Cord}{\QCI},
\author{Chan U Lei}{\QCI},
\author{Joseph O. Yuan}{\QCI},
\author{Shantanu O. Mundhada}{\QCI},
\author{Kevin S. Chou}{\QCI},
\author{S. Harvey Moseley, Jr.}{\QCI},
\author{Robert J. Schoelkopf}{\QCI,\YalePhysics,\YQI}

\bigskip

\xQCI
\xYalePhysics
\xYaleQuantumInstitute
\xMicrosoft
\xWeizmann
\xEqualContribution

}

\tocless\section{Author contributions}
J.D.T. and S.O.M. contributed to the proposal and concept of the SWS gate. 
S.O.M. led the design of the dual-rail hardware architecture. 
J.O.Y led hardware deployment, experimental setup, and system validation.
B.C. led the device fabrication.
C.U.L. led the cavity design, preparation, and testing.
K.S.C. led two-qubit gate testing, measurements, and analysis.
N.M., T.N., P.L., B.H.E., A.Agrawal, A.N., K.S.C, J.O.Y., and J.D.T. contributed to data acquisition and data analysis.
T.N., N.M., P.L., A. Agrawal, K.S.C., and J.D.T. contributed to the SWS gate calibration procedure.
J.D.T. and K.S.C. contributed to numerical simulations of the SWS gate.
P.L., N.M., J.O.Y., T.N., A. Agrawal, B.H.E., G.L., K.S.C., and N.K. contributed to system bring-up and calibration.
S.O.M., R.C., S.M.F., T.Z.C., R.D.L., H.M., D.K.W., and J.O.Y contributed to quantum hardware design and simulation.
E.L. B.B., M.C., C.F., M.R. contributed to mechanical design.
S.O.M. and R.C. contributed to coupler design and flux-pump delivery.
C.U.L., C.K., J.O.Y., B.H.E., and N.K. contributed to cavity preparation and testing.
J.S., A.B., F.D., H.H., D.V.S. contributed to device assembly and integration.
D.D., J.O.Y., M.Maxwell, and J.S. contributed to laboratory infrastructure.
S.S, M.Mirzaei, D.S., B.G., B.C., H.L., and L.Y. performed device fabrication. 
M.Maxwell, E.R., A.C., S.R.D., R.J., R.P., and O.R designed, tested, and built the RF control system. 
J.O.Y., S.O.M., G. L., N.K. and R.A.S. contributed to RF wiring and implementation of the readout chain.
J.C.C., P.L., N.M., D.K.W, K.S.C., and T.K. contributed to data analysis of randomized benchmarking results and supporting simulations.
J.D.T., S.O.M., A.N, N.M., T.N., K.S.C., H.M., T.Z.C., N.K., and T.S. contributed to an earlier version of the SWS gate.
M.S., S.W., D.Sweeney, A.Anderson, W.B., J.C., L.C., K.K., T.K., A.M., L.M-K., D.Szperka, I-W.S., Z.Y. contributed to software infrastructure and calibration tools used in this project.
K.S.C, N.M., T.N., J.D.T., and R.J.S. wrote the manuscript.
R.J.S., S.H.M., J.A., S.H., O.B. C-L.T. managed the project. 

\bigskip
\tocless\section{Materials and Correspondence}
Correspondence and requests for materials should be addressed to Nitish Mehta (mehta@quantumcircuits.com), Kevin Chou (chou@quantumcircuits.com), and Robert J. Schoelkopf (robert.schoelkopf@yale.edu)

\tocless\section{Competing interests statement}
R.J.S. is a founder and shareholder of Quantum Circuits, Inc (QCI).
Authors affiliated with QCI have financial interest in the company.

\bibliographystyle{naturemag}
\let\oldaddcontentsline\addcontentsline
\renewcommand{\addcontentsline}[3]{}
\bibliography{bib}
\let\addcontentsline\oldaddcontentsline

\clearpage
\appendix

\onecolumngrid
\begin{center}
{\large \textbf{Supplemental information for ``Bias-preserving and error-detectable entangling operations in a superconducting dual-rail system"} \par}
\end{center}
\twocolumngrid

\tableofcontents

\section{Single-qubit characterization}
\label{app:1q_characterization}
\subsection{1Q gate characterization}

\begin{table*}
    \centering
    \begin{tabular}{| c | c | c | c || c | c | }
    \hline
    & Description & Parameter & Units & $\text{Control qubit}$ & $\text{Target qubit}$ \\ 
     \hline\hline
     \rule{0pt}{2.6ex}
    System & Cavity $b_1$-Coupler $c$ cross-Kerr & $\chi_{\mathrm{bc}}/2\pi$ & MHz &  \multicolumn{2}{c|}{-1.51}  \\ [1mm] 
    & Cavity $a_2$-Coupler $c$ cross-Kerr & $\chi_{\mathrm{ac}}/2\pi$ & MHz  & \multicolumn{2}{c|}{-1.26} \\[1mm] 
     & Cavity $a_2$-Cavity $b_1$ cross-Kerr& $\chi_{\mathrm{ab}}/2\pi$ & kHz & \multicolumn{2}{c|}{-6.64} \\[1mm]
     & Cavity $a_2$-Coupler $c$ photon exchange rate & $g_{\mathrm{ac}}/2\pi$ & MHz  & \multicolumn{2}{c|}{4.23}  \\[1mm]  
    \hline
      \rule{0pt}{2.6ex}
    Cavity  & Relaxation time & $T_{1}^{\mathrm{a (b)}}$ &$\upmu$s& 231, 411 & 652, 342\\
    [1mm]  
      & Decoherence time, Ramsey & $T_{\mathrm{2R}}^{\mathrm{a (b)}}$ &$\upmu$s& 433, 638 & 1001, 454\\[1mm]
    & Resonance frequency & $\omega_{\mathrm{a (b)}}/2\pi$ & GHz & 7.03, 6.75 & 6.56, 7.21 \\[1mm]
    \hline
    \rule{0pt}{2.6ex}
    Entangling coupler & Relaxation time & $T_{1}^{\mathrm{c}}$ &$\upmu$s & \multicolumn{2}{c|}{70}  \\[1mm]
     & Dephasing time, Ramsey  & $T_{\phi}^{c}$ &$\mu$s& \multicolumn{2}{c|}{327}   \\[1mm]
      & Dephasing time, Echo  & $T_{\phi}^{c}$ &$\mu$s& \multicolumn{2}{c|}{1001}   \\[1mm]
      \hline
    \end{tabular}
    \caption{Measured Hamiltonian and coherence properties.}
    \label{tab:system_properties}
\end{table*}

\begin{table*}
    \begin{tabular}{| c | c || c | c |}
    \hline
    Property & Parameter & Control qubit & Target qubit \\
     \hline\hline
    SPAM [1 round] & Average misassignment  & $2 \times 10^{-4}$ & $2.3 \times 10^{-4}$ \\
     & Average leakage detection error & $3.46 \times 10^{-3}$ & $3.90 \times 10^{-3}$ \\
     & Average erasure assignment & $6.79 \times 10^{-2}$ & $6.43 \times 10^{-2}$\\
     \hline
    SPAM [2 rounds] & Average misassignment  & $7 \times 10^{-6}$& $1.2 \times 10^{-5}$ \\
     & Average leakage detection error & $1.5 \times 10^{-4}$ & $1.5 \times 10^{-4}$  \\
     & Average erasure assignment & $1.81 \times 10^{-1}$ & $1.41 \times 10^{-1}$\\
     \hline
    Coherences & Ramsey time  & $3.1 \: \mathrm{ms}$  & $1.5 \: \mathrm{ms}$\\
     & Echo time & $4.0 \: \mathrm{ms}$ & $4.8 \: \mathrm{ms}$\\
     & Bit-flip & $520 \:\mathrm{ms}$  &  $1100 \: \mathrm{ms}$ \\
     \hline
    Single-qubit gates & Individual gate error [iRB]  & $9.0 \times 10^{-5}$ & $1.08 \times 10^{-5}$ \\
     & Simultaneous gate error [iRB] & $1.01 \times 10^{-4}$ & $1.36 \times 10^{-4}$ \\
     & Erasure per $\sqrt{X}$ gate & $8.56 \times 10^{-4}$ & $5.92 \times 10^{-4}$ \\
     & Total gate duration & $208 \:\mathrm{ns}$& $136\: \mathrm{ns}$ \\
     & Beamsplitter rate & $1.78 \: \mathrm{MHz}$ & $2.55 \: \mathrm{MHz}$ \\
     \hline
    \end{tabular}
    \caption{Measured dual-rail qubit properties.}
    \label{tab:DR_coherences}
\end{table*}

\begin{figure}[ht]
\centering
\includegraphics[width=\columnwidth]{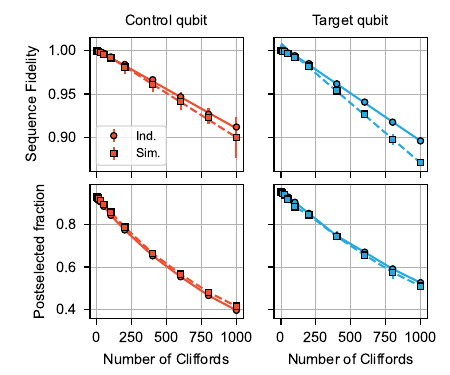}
\caption{
    \label{fig_1q_rb}
    \textbf{Single-qubit gate benchmarking}
    Individual (circles) and simultaneous (square) RB on the control qubit (left) and target qubit (right). Control qubit individual and simultaneous RB are measured to be $9.0(2)\times 10^{-5}$ and $1.01(1)\times 10^{-4}$, respectively. Target qubit individual and simultaneous RB are measured to be $1.08(1)\times 10^{-4}$ and $1.36(1)\times 10^{-4}$. From postselected fraction we quantify erasure rates of $0.0856(5)\%$ and  $0.0592(6)\%$ for control and target qubit, respectively. Results show minimal change in gate fidelity when operating the two qubit simultaneously and that there are low cross-talk errors in this system. 
}
\end{figure}

Single-qubit gates are realized by four wave mixing parametric beamsplitter-type interactions implemented by a SQUID-transmon coupler \cite{lu_high-fidelity_2023}. We apply a flux drive to the SQUID coupler at half the frequency detuning between the two cavities ($a$ and $b$) of a dual-rail qubit to enable a parametric interaction via the Hamiltonian $\mathcal{H}/\hbar = \frac{g_{\mathrm{bs}}}{2}\left( \hat{a}^\dagger \hat{b} + \hat{a} \hat{b}^\dagger\right)$.
To characterize single-qubit gates, we perform both individual and simultaneous RB. For these measurements we use a gateset consisting of $\{X_{\frac{\pi}{2}}, Z, Z_{\pm \frac{\pi}{2}}\}$ from which we generate the 24 single-qubit Cliffords. For each Clifford sequence length we average over 20 different randomized sequences. 
Performing RB individually on control and target qubit, we extract a gate fidelity in excess of $>99.98\%$ with a small decrease in postselected fraction of $< 0.1\%$ per Clifford detected by our end-of-the-line measurement 
(see Fig.~\ref{fig_1q_rb}). When operated simultaneously, the control and target qubit fidelities are observed to modestly decrease by $0.001\%$ and $0.002\%$, respectively for the control and target qubits. We attribute this effect to a small measurable cross-Kerr between two dual-rail qubits ($a_2$ and $b_1$ cavities) of  $\mathrm{6.64 \: kHz}$. 
Compared to our single-qubit beamsplitter rates of $1.78 \mathrm{ \: MHz \:}   (2.55 \mathrm{\: MHz })$ for the control (target) qubit we estimate simultaneous operation infidelities to be limited by $(\chi_{ab}/g_{aa (bb)})^2 \sim 1.3 \times 10^{-5}$.

\subsection{SPAM}
Dual-rail measurement are implemented by performing simultaneous single-shot cavity measurements that determine whether a photon is present in each cavity individually, extracting whether the system is in $\ket{01}$ or $\ket{10}$. This measurement is designed to also have an important property: it can detect for both cavity photon loss and also can flag hardware errors during the measurement protocol, like decoherence or transmon readout misassignments, during the measurement itself. Detailed characterization of dual-rail SPAM are shown in Table~\ref{tab:DR_coherences}. In addition, this measurement can be repeated for multiple rounds and correlating the results can further suppress misassignment errors. This postselection capability comes at the cost of a first-order decrease in postselected fraction, however, we benefit from a exponential decrease in missassignments, which is highly beneficial for characterization of our CZ gate. Additional details can be found in ref~\cite{chou_superconducting_2024}. We benchmark performance for state preparation and measurement (SPAM) and extract logical misassignment errors of $0.023\% $ and $0.020\% $ for the target and control dual-rail qubit respectively. The overall SPAM erasure errors, defined to be assigned erasure outcomes when attempting to prepare in either $\ket{01}$ or $\ket{10}$, are measured to be $6.79\%$ and $6.43\%$ for the target and control qubit. Finally we benchmark the leakage detection properties by initializing the dual-rail qubit in the $\ket{00}$ and we observe erroneous codespace assignments to be $0.39\%$ and $0.34\%$ for the target and control qubit.

\section{Two-qubit gate calibration}\label{2Qcalib}

\begin{figure*}[ht]
\centering
\includegraphics[width=\textwidth]{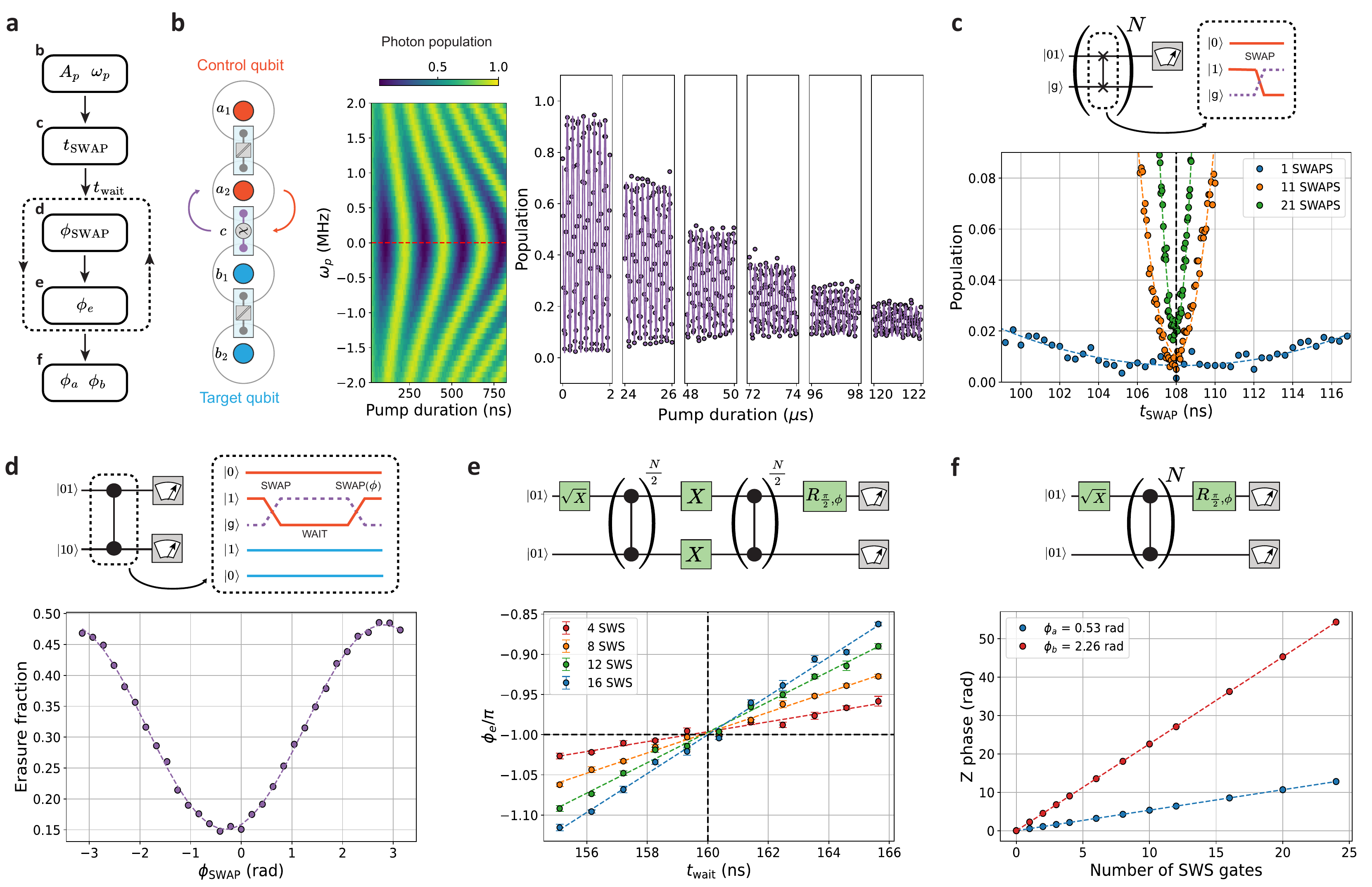}
\caption{
    \label{fig_cz_calibration}
    \textbf{Calibration of CZ gate.}
    \textbf{a} Flowchart of the CZ gate calibration procedure.
    \textbf {b} Left: Schematic illustration of swap between the control qubit cavity $a_2$ (red) and coupler $c$ (purple). Right: Photon population in the control qubit cavity as a function of parametric pump frequency $\omega_p$ and duration of the pump, and additionally the stroboscopic evolution at the resonant pump frequency marked by a red dashed line.
    {\textbf{c}-\textbf{f.} Each Fig. shows the pulse sequence and the corresponding calibration result.}
    \textbf {c.} Calibration of duration of cavity-coupler swap by measuring the population in the control qubit cavity after odd numbers of swap.
    \textbf {d.} Calibration of the phase of rotation for the second swap operation in the CZ gate by minimizing the erasure fraction. 
    \textbf{e.} Calibration of the delay time of the CZ gate by measuring the entangling phase per CZ gate, which should corresponds to $\pi$ for the calibrated delay time marked by dashed line.
    \textbf{f.} Calibration of single-qubit Z rotations per CZ gate. The Ramsey-like protocol should be performed on either the control or target qubit to measure Z rotation of each qubit.
}
\end{figure*}

In this section, we discuss the calibration of the CZ gate in detail. The calibration procedure is illustrated in the flowchart in Fig.~\ref{fig_cz_calibration}a, which starts with calibrating the photon swap operation between the control qubit cavity $a_2$ and the coupler $c$, referred to as the cavity-coupler swap. The underlying interaction for this cavity-coupler swap is identical to that used for the single-qubit gate. However, in this case, the parametric flux pump is applied at half the frequency difference of the Stark shifted coupler $c$ and cavity $a_2$. A coherent transfer of a single photon between the cavity and the coupler is shown in Fig.~\ref{fig_cz_calibration}b where we also present the stroboscopic time evolution of the photon population in the cavity $P_{\rm{cav}}$ at the resonant pump frequency, which follows
\begin{equation}
P_{\rm{cav}} (t)  = \frac{1}{2} e^{- {\kappa t}} \left(1+e^{-\kappa_\phi t} {\rm{cos}}(g_{ac} t) \right).
\end{equation}
Here, $g_{ac}$ represents the rate of the cavity-coupler swap, while $\kappa$ and $\kappa_\phi$ denote the photon decay rate and the dephasing rate when the pump is turned on.
As the first step in calibrating the cavity coupler swap operation, we find the amplitude $A_p$ that maximizes the metric $g_{ac}/\kappa_{\phi}$, which leads to the highest fidelity of the cavity-coupler swap operation when you postselect for photon loss. Following this, we calibrate the duration of the swap pulse $t_{\mathrm{SWAP}}$ by repeatedly applying the pulse after starting with a photon in cavity $a_2$ and minimizing the population in the cavity $a_2$ at the end of odd number of swap operations 
(see Fig.~\ref{fig_cz_calibration}c).
When the swap operation is used in the SWS gate sequence, an additional parameter needs to be specified which is the relative phase difference between the first and the second swap operations. To understand why this is important, we first note that at the end of the CZ gate we want all the population in cavity $a_2$ to return to its original value irrespective of which two qubit state we start in. This condition is trivially satisfied for starting states $\ket{10g01}, \ket{10g10}$ and $\ket{01g01}$ since in these cases either no transfer of population takes place between cavity $a_2$ and coupler $c$ or an on-resonant swap takes place, either way the phase of the second swap operation is irrelevant. The only non-trivial case occurs when we start in the state $\ket{01g10}$, in which case we perform a swap operation detuned by $\chi_{bc}$ (see Fig.~\ref{fig:bloch_traj}). In this case we can close the Bloch sphere trajectory by choosing the correct phase of the second swap operation in the gate. To calibrate this parameter we initialize our system in the state $\ket{01g10}$ and apply the SWS gate sequence while sweeping the phase of the second swap pulse and measure the total erasure fraction of the circuit. We choose the swap back phase $\phi_{\mathrm{swap}}$ that minimizes the erasure fraction (See Fig.~\ref{fig_cz_calibration}d). Since our gate is excitation-preserving, every time we leave population in the coupler mode $c$ we will find the control qubit in the vacuum state. Therefore, minimizing the erasure fraction is equivalent to maximizing the population in cavity $a_2$ at the end of the gate. It is important to note that as the delay time varies, the swap back phase should be calibrated and adjusted accordingly prior to measuring the corresponding entangling phase $\phi_e$ as illustrated in the flowchart, until the delay time is accurately calibrated.

With the cavity-coupler swap time and the swap back phase for a given delay time $t_{\mathrm{wait}}$ calibrated, the unitary of the SWS gate can be written as the following, 
\begin{equation}
    U_{\rm{SWS}} = \begin{pmatrix}
1 &  0&  0&  0\\
0 & e^{i\phi_a} & 0 & 0  \\
0 & 0 & e^{i\phi_b} & 0 \\
0 & 0 & 0 & e^{i(\phi_a +\phi_b+\phi_e)} 
\end{pmatrix}
\end{equation}
where $\phi_e$ is the entangling phase, $\phi_a$ and $\phi_b$ are respective $Z$-phase on each qubit accumulated during the gate. Next, to realize CZ unitary, we calibrate the wait time $t_{\mathrm{wait}}$ such that $\phi_e = \pi$. For cavity-coupler swaps with instantaneous pulse rise times and zero cavity-cavity cross-Kerr $\chi_{ab}$ between the cavities $a_2$ and $b_1$, the wait time can be analytically calculated (see Eq. \ref{twait}). However, in the case of non-zero pulse rise time and $\chi_{ab}$, we need to experimentally determine the wait time. To measure $\phi_e$ for a given wait time, we perform a Ramsey-like sequence \cite{google_mead} on either the control or the target qubit . In this protocol, we start with both qubits in the state $\ket{0_L}$ and apply a $\sqrt X$ gate to one of the qubits to prepare a superposition state $\frac{1}{\sqrt{2}} (|0_L 0_L\rangle -i |1_L 0_L\rangle )$. Following that we repeatedly apply $N$ CZ gates interrupted by $X$ gates on both qubits in the middle. This protocol then creates the state $\frac{1}{\sqrt{2}} (e^{i N \phi_e/2}|1_L 1_L\rangle -i |0_L 1_L\rangle )$ where the phase $N\phi_e/2$ is accumulated as a relative phase shift between the states and the phases $\phi_a$ and $\phi_b$ are factored out as global phases. The entangling phase $\phi_e$ measured as a function of wait time $t_{\mathrm{wait}}$ and number of repeats is shown in Fig.~\ref{fig_cz_calibration}e from which we find the optimal wait time that leads to $\phi_e = \pi$.

Once the delay time is precisely calibrated, the last two parameters to be calibrated are the $Z$-phase rotation for the control and target qubit $\phi_a$ and $\phi_b$. These can also be measured via Ramsey-like protocol on the control and target qubit in similar manner as the entangling phase $\phi_e$. The only difference is that we do not perform $X$ gate in the middle of the sequence as the echo pulses will convert the phases $\phi_a$ and $\phi_b$ that we are interested in measuring into in to global phases. Fig.~\ref{fig_cz_calibration}f shows these local Z-phases as a function of number of repeated CZ gates for  the wait time corresponding to entangling phase $\phi_e = \pi$ where the extracted slope of each trace corresponds to the phases $\phi_a$ and $\phi_b$ accumulated per gate. 

\section{CZ gate on-off ratio}

\begin{figure}[ht]
\centering
\includegraphics[width=3.0in]{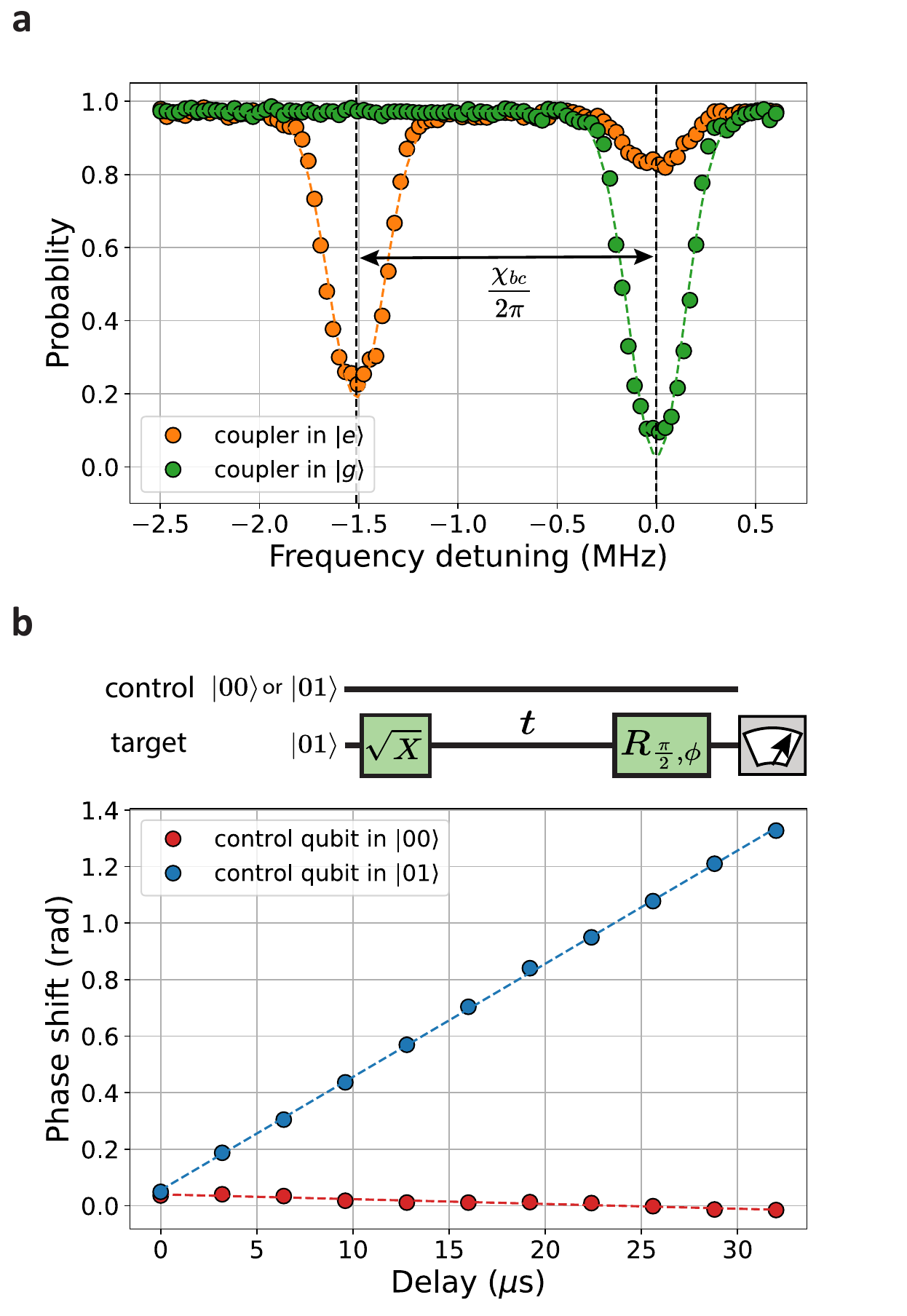}
\caption{
    \label{fig_chi}
    \textbf{Measuring interaction strength for the CZ gate.} 
    \textbf {a.} Spectroscopy on the target qubit cavity with the coupler either in the ground (green) or excited state (orange). 
    \textbf{b.} Circuit diagram for the Ramsey sequence on the target dual-rail and the associated phase shift as a function of delay time with the control qubit cavity either empty (red) or occupied with a photon (blue).
}
\end{figure}

As described in the main text, the construction of our CZ gate is based on the dispersive interaction between the coupler and the target dual-rail, $\chi_{bc} \hat{b}_1^\dagger \hat{b}_1 \hat{c}^\dagger \hat{c}$, where $c$ and $b_1$ denote the coupler and target qubit cavity connected to the coupler, respectively. Therefore, once the interaction is activated, the entangling phase accumulates at a rate $\chi_{bc}$, which can be readily measured by performing spectroscopy on the target qubit cavity $b_1$. Fig.~\ref{fig_chi}a shows the results with the coupler prepared in either the ground or excited state with a choice of applying the cavity-coupler swap prior to spectroscopy. In the case where the coupler is in the excited state, the decay of the population in the coupler from the excited to the ground state leads to an additional dip at zero frequency detuning. By comparing the resonance frequency of the target qubit cavity $b_1$  depending on the state of the coupler, we can extract $\chi_{bc}/{2\pi} \sim 1.51$ MHz as marked in the figure. On the other hand, the dispersive interaction $\chi_{ab} \hat{a}_2^\dagger \hat{a}_2 \hat{b}_1^\dagger \hat{b}_1$ between the control qubity cavity $a_2$ and target qubit cavity $b_1$ leads to {\textit{unwanted}} entanglement between the two dual-rail qubits when they are idle. In order to accurately extract $\chi_{ab}$, we apply the Ramsey sequence on the target qubit with a given delay time $t_{\mathrm{wait}}$. We measure the probability of being in $|01\rangle$ after the sequence which oscillates as a function of the phase of the second $\pi/2$ pulse. Fig.~\ref{fig_chi}b shows the associated phase shift in the oscillation for various delay times, with the control qubit cavity either empty or occupied with a photon during the Ramsey sequence. By taking the difference in the slopes of the linear fit to the respective results, we can extract $\chi_{ab}/{2\pi} \sim 6.64$ kHz, which leads to an on-off ratio of the CZ gate $\sim 230$. 

\section{Inferring gate fidelities}
\label{app:computing_fidelities}
In this section we provide additional detail for how we estimate CZ gate fidelity in the short-depth limit.

Following the treatment in Ref.~\cite{magesan_efficient_2012}, consider the decay in RB survival probability as a function of the number of gates $N$, $F = A p^N + B$, where $p$ corresponds to the depolarizing parameter. 
We first substitute $p = 1-x$, and expand to linear order in $N$
\begin{align*}
    F &= A (1-x)^N + B = A (1-xN) + B \\
    &= -Ax \cdot N + A + B
\end{align*} 
We perform a linear fit to our data and extract a slope $m = -Ax$ and can see that the depolarizing parameter $p$ can be computed as $p = 1 - |m| / A$. From this analysis, we see that the value of $p$ depends on the amplitude and offset of the survival probability. 

The average CZ gate error is computed from interleaved RB as:
\begin{align}
    r_\mathrm{CZ} &= \frac{(d-1) \left(1 - p_\mathrm{i} / p_{r}\right)}{d} 
\end{align}
with $d = 2^{N_q} = 4$, $p_{i,r}$ are the depolarizing parameter for interleaved and reference RB, respectively. 

For two-qubit RB we expect $A = \frac{3}{4}$ and $B = \frac{1}{4}$, and thus
\begin{equation}
    r_\mathrm{CZ} = \frac{3}{4} \left(1 - \frac{1 - \frac{4}{3}|m_i|}{1 - \frac{4}{3}|m_r|} \right) = \frac{3}{4} \frac{|m_i| - |m_r|}{\frac{3}{4} - |m_r|},
\end{equation}
where $m_{i,r}$ are the extracted slopes for the interleaved RB and reference RB data, respectively.

For one-qubit RB, $A = B = \frac{1}{2}$ we extract the average gate error as 
\begin{equation}
    r_{1q} = \frac{d-1}{d} (1 - p) = \frac{1}{2} \frac{A}{|m|} = \frac{1}{4|m|}
\end{equation}

We also consider extracting gate fidelity from decay of Bell state fidelity. For our CZ gate in the limit of large number of repeated gates $N$, the state fidelity will saturate to $B=1/4$. Using $A = 3/4$, then the we find the CZ gate fidelity $p = 1 - \frac{4|m|}{3}$.

\section{Evaluating the accuracy of IRB with postselection}

\begin{figure}
    \centering
    \includegraphics[]{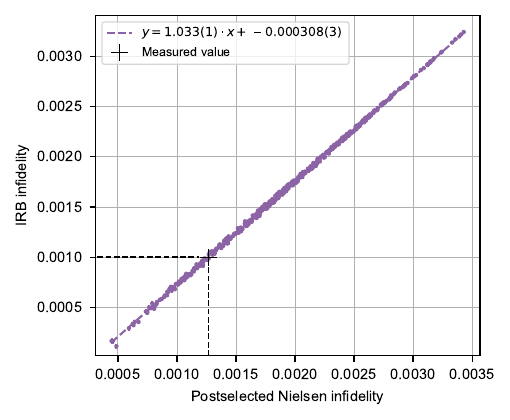}
    \caption{$\mathrm{CZ}$ gate infidelity inferred from simulated IRB experiments vs. the infidelity computed from Eq.~\ref{eq:nielsen_fid}. 500 random dephasing and leakage rates are chosen uniformly distributed with $p^{\text{control}}_Z, \,p^{\text{target}}_Z\in[1\times10^{-4}, 2\times10^{-3}]$, $p_{ZZ}\in[5\times10^{-5}, 2\times10^{-4}]$, and $p^{\text{control}}_{\text{leakage}},\,p^{\text{target}}_{\text{leakage}}\in[1\times10^{-4}, 4\times10^{-3}]$. Each IRB fidelity simulation uses the exact same random gate sequence as the measurement in Fig.~\ref{fig2}. The solid lines guide the eye to convert our measured $\mathrm{CZ}$ infidelity to the infidelity when error detection immediately follows the gate.
    }
    \label{fig:IRB_accuracy}
\end{figure}

Appendix \ref{app:computing_fidelities} utilizes IRB to estimate $\mathrm{CZ}$ gate fidelities at short depths with a linear approximation of the decay exponential. Here, we demonstrate that the experimental results are a reliable estimator for the true underlying gate errors. The addition of error detection to RB is novel and, to our knowledge, only utilized here and in neutral atom platforms \cite{ma_high-fidelity_2023}. These error detection measurements are high-fidelity \cite{chou_superconducting_2024}, but leakage misassignment errors do occur at $10^{-4}$ rates. Additionally, the infidelity of our $\mathrm{CZ}$ gate exceeds those of single qubit rotations by more than one order of magnitude. This produces an asymmetry between some Cliffords and the interleaved $\mathrm{CZ}$ gate that can negatively impact the $\mathrm{CZ}$ fidelity estimate \cite{epsteinInvestigatingLimitsRandomized2014}. Between the addition of error detection, gate asymmetry, and limited circuit depth, it is valid to worry that IRB 
 may not provide an accurate estimate of $\mathrm{CZ}$ gate fidelity. In this section, we define a parametrized channel for our $\mathrm{CZ}$ gate, add postselection to the standard definition of average gate fidelity \cite{nielsenSimpleFormulaAverage2002}, and simulate IRB under different $\mathrm{CZ}$ error rates to evaluate its estimates of the postselected gate fidelity.

First, we discuss a simplified leakage model for a DR qubit. For a single DR, we represent the logical and leakage states with qutrit states $\ket{0}=\ket{0_L}$, $\ket{1}=\ket{1_L}$, $\ket{2}=\ket{*_L}$. 
Thus, for each $x\in\{\mathrm{target},\;\mathrm{control} \}$ in a two DR system, the leakage channel is:
\begin{align}
    \Lambda^{x}_{\text{leakage}}(\rho)&= (1-p^x_{\text{leakage}})\rho+p^x_{\text{leakage}}K^x\rho (K^x)^\dagger
\end{align}
with $K^x=\ket{*_L}_{x\;x}\bra{0_L}+\ket{*_L}_{x\;x}\bra{1_L}+\ket{*_L}_{x\;x}\bra{*_L}$.

Detectable leakage errors form the dominant component of the $\mathrm{CZ}$ error channel, but residual dephasing arises from intrinsic sources and control qubit leakage during gates. To form the dephasing error channel, note that dephasing in the computational subspace $\{\ket{0_L},
\ket{1_L}\}$ is generated by the $\sigma_z$-like Gell-Mann matrix $\lambda_3$  \cite{gell-mannSymmetriesBaryonsMesons1962} with non-zero diagonal $(1,-1,0)$. The resulting dephasing Kraus operator on a single qutrit is $K_3=\exp{(-i\pi\lambda_3/2)}$. This produces the two qutrit dephasing channel:
\begin{align}
    \Lambda_{\text{dephas.}}(\rho)&=(1-p^{\text{control}}_Z-p^{\text{target}}_Z-p_{ZZ})\rho \\
    &\;\;+p^{\text{control}}_Z K_3^{\text{control}}\rho (K_3^{\text{control}})^\dagger \nonumber \\
    &\;\;+p^{\text{target}}_Z K_3^{\text{target}}\rho (K_3^{\text{target}})^\dagger \nonumber \\
    &\;\;+p_{ZZ}(K_3^{\text{control}}\otimes K_3^{\text{target}})\rho(K_3^{\text{control}}\otimes K_3^{\text{target}})^\dagger. \nonumber
\end{align}
The $Z$ subscripts serve as a reminder that dephasing occurs in the qubit subspace.

Finally, we can write a simple $\mathrm{CZ}$ channel $\mathcal{E}$ as the composition of leakage and dephasing channels
\begin{align}
    \mathcal{E}(\rho)&=\Lambda^{\text{control}}_{\text{leakage}}\circ\Lambda^{\text{target}}_{\text{leakage}}\circ\Lambda_{\text{dephase}}(\mathrm{CZ}\rho\mathrm{CZ}^\dagger).
\end{align}
This channel captures target qubit dephasing resulting from control qubit leakage during the $\mathrm{CZ}$ gate, albeit more simply than the detailed picture presented in Appendix~\ref{app:leakage_prop}. The rates parametrizing each individual error channel are derived from the error-per-gate results in Fig.~\ref{fig3}.

Our RB circuits only include leakage detection at the end of a circuit, unlike other approaches that interleave checks with batches of gates \cite{ma_high-fidelity_2023,Levinedualrail2024}. Here we explain why this technique is valid for our system. DR qubit seepage via cavity photon heating occurs at a rate at least 1000 times slower than leakage~\cite{chou_superconducting_2024}, obviating the need for mid-circuit checks here or other more sophisticated techniques~\cite{woodQuantificationCharacterizationLeakage2018a,chenRandomizedBenchmarkingLeakage2025}. This means that the fidelity of a standalone gate immediately followed by leakage detection should be nearly equivalent to the fidelity of a gate with delayed leakage detection. An RB experiment with postselection should then infer the fidelity of a postselected gate, up to misassignment errors in the leakage detection. This motivates a modification to the standard entanglement fidelity definition found in \cite{nielsenSimpleFormulaAverage2002}:
\begin{align}
    \label{eq:nielsen_fid}
    F_{e}(\mathcal{E},\mathrm{CZ})=\sum_{j,k}\frac{\alpha_{jk}\mathrm{Tr}(\mathrm{CZ}U_{j}\mathrm{CZ}(M\mathcal{E}(\rho_{k})M^{\dagger}))}{d^3\mathrm{Tr}(M\mathcal{E}(\rho_{k})M^{\dagger})}.
\end{align}
The $U_j=\sum_k\alpha_{jk}\rho_k$ form a unitary basis on the two qubit space \cite{nielsenSimpleFormulaAverage2002, klappeneckerStabilizerCodesNice2001} with pure state representation $\rho_{k}$. Using a state basis here is essential for postselection after propagation through the $\mathrm{CZ}$ channel $\mathcal{E}$ as the $U_j$ can be traceless. The postselection measurement operator $M$ projects into the two-qubit codespace, with misassignments characterized in Table~\ref{tab:DR_coherences} that appropriately penalize the gate fidelity \cite{chou_superconducting_2024}. The denominator in Eq. \ref{eq:nielsen_fid} renormalizes the fidelity contribution of each $\rho_{k}$ by its postselected fraction. This is similar to the approach taken in \cite{weissQuantumRandomAccess2024}, where the entanglement fidelity is divided by the average postselected fraction with the intent to preserve channel linearity.

Now that we can compute the postselected fidelity of our $\mathrm{CZ}$ channel, we compare this to the fidelity inferred from IRB simulations across a range of $\mathrm{CZ}$ error rates. We follow a standard technique used in \cite{nielsenSimpleFormulaAverage2002, magesan_efficient_2012, epsteinInvestigatingLimitsRandomized2014} using the superoperator formalism to propagate a density matrix through a random sequence of gates. There are two more gates required to generate the Clifford group: $X_{\frac{\pi}{2}}$ and virtual-$Z$. We compute the superoperator implementing both individual and simultaneous $X_{\frac{\pi}{2}}$ operations using the coherences and gate durations in Tables~\ref{tab:system_properties} and \ref{tab:DR_coherences}. It is important to include the residual $a_1-b_2$ cross-Kerr to capture unitary gate errors.  Virtual-$Z$ gates are perfect unitaries in the simulation.

Having described our model for all operations, we proceed to simulate IRB and extract two-qubit gate fidelities. 
All of our simulations implement the same random gate sequence used for the measurement in Fig.~\ref{fig2}. Due to finite sampling of random sequences (20 at each depth), we expect noise in the inferred infidelity particular to each sequence. Using error rates centered around estimated values from Fig.~\ref{fig3}, we find that IRB systematically underestimates infidelities across the range, including by $25\%$ at our measured value (marked on Fig.~\ref{fig:IRB_accuracy}). In absolute terms, this only adjusts the infidelity from $0.1\%$ up to $0.13\%$, confirming a high-fidelity gate. Other investigations have found comparable fractional errors at similarly high gate fidelities, albeit under maximally deleterious unitary errors \cite{epsteinInvestigatingLimitsRandomized2014}. The underestimate persists even when using perfect leakage detection measurements in the simulation, suggesting that the bias may be a product of this particular set of random gates. Simulations of other random sequences produce different slopes and offsets such as $0.984(1)$ and $-0.000150(2)$, respectively, supporting this claim.

\section{Leakage propagation}
\label{app:leakage_prop}
In this section we will provide more details supporting our investigation of the leakage propagation properties of our CZ gate. We consider leakage before and during the CZ gate and derive error models for both cases. 

\begin{table}
    \centering
    \begin{tabular} { c | c  }
        \hline \hline
        Control qubit action & Effect on target qubit \\
        \midrule
        start in $\ket{10}$ & I \\
        start in $\ket{01}$ & Z \\
        start in leakage state $\ket{00}$ & I \\
        start in $\ket{10}$ and leaks during the gate & I \\        
        start in $\ket{01}$ and leaks during the gate & $Z(\theta), \theta \in [0,\pi ]$
        \\
        \hline \hline
    \end{tabular}
    \caption{Leakage propagation properties for CZ gate}
    \label{tab:leakage_propogation}
\end{table}

\subsection{Leakage before a CZ gate}
We first consider the case when leakage occurs before the CZ gate and either the control or target qubit starts in the ground state, $\ket{00}$. In this situation the dispersive interaction between coupler and target qubit is never switched on; this interaction requires one excitation in the coupler and one in the target qubit, but with a leakage there is only one total excitation between the two qubits. As such the overall process when we suffer leakage before the CZ gate is an overall identity process on the other qubit.

\subsection{Leakage during a CZ gate}
We now derive the leakage error model for the CZ gate. For this model to be valid, we require that only two of the four cavities interact during the gate, and the CZ gate is realized using a \texttt{CPHASE}-type interaction, that is, only one of the four two-qubit basis states gains an entangling phase at some steady rate. This can be achieved with the following general interaction Hamiltonian
\begin{equation}
    \mathcal{H}_\mathrm{ent} / \hbar = \chi_\mathrm{ent} \outerproduct{1_L1_L}{1_L1_L} \equiv \chi_{\mathrm{ent}} 
 \mathrm{CZ} 
\end{equation}
The CZ gate is then implemented by unitary evolution of this Hamiltonian for a time $t_\mathrm{CZ} = \pi / \chi_\mathrm{ent}$. 

A leakage event occurs at some random time during the gate, resulting in first a CZ 
 interaction of some random phase, followed by leakage to $\ket{00}$ that turns off the interaction. The average channel that results from this CZ of some unknown angle $\phi$ can be computed by evaluating the following integral
\begin{align}
    \Lambda_\mathrm{leakage}(\rho)&= \frac{1}{\pi} \int_0^\pi \mathrm{CZ}(\phi)~ \rho~ \mathrm{CZ}(-\phi) d\phi,
\end{align}
Where $\Lambda_\mathrm{leakage}(\rho)$ is the two-qubit error channel applied prior to the leakage jump operator.
To solve this integral, we make use of the following decomposition 
\begin{align}
    \mathrm{CZ}(\phi)
    &= ZZ(\phi/2) Z_a(-\phi/2) Z_b(-\phi/2) \\
    &= x(\theta) I + y(\theta) \left(Z_a + Z_b - ZZ \right),
\end{align}
where $ZZ(\phi) = Z_a(\phi) \otimes Z_b(\phi)$,  $x(\theta)=\cos^3(\theta/2)+i\sin^3(\theta/2)$, and $y(\theta)=ie^{i\theta/2}\sin(\theta/2)\cos(\theta/2)$. 
We also make use of the relation $\mathrm{CZ} = \mathrm{CZ}(\pi) = \frac{1}{2} \left(I + Z_\mathrm{a} + Z_\mathrm{b} + ZZ \right)$.

After collecting and integrating these trigonometric functions we arrive at
\begin{equation}
    \label{eq:leakage_process}
    \Lambda_\mathrm{leakage}(\rho) = 
    \frac{1}{2} \left(\mathds{1}\rho\mathds{1} + \mathrm{CZ}\rho \mathrm{CZ} \right) + \frac{4i}{3\pi} \left(\mathrm{CZ} \rho \mathds{1} - \mathds{1}\rho \mathrm{CZ} \right),
    \end{equation}
where the channel splits into the first two ``diagonal" terms and the ``off-diagonal" imaginary terms that result from the direction through which we acquire entangling phase. These off-diagonal elements are also visible in Fig.~\ref{fig4}e in the imaginary part of the $\chi$-error matrix for $N=1$ repeats. They deviate in magnitude from $4/3\pi$ due to a significant fraction of the leakage occurring during state preparation and measurement, rather than the CZ gate itself.

In an error correction context, measuring stabilizers will ``digitize'' our quantum errors and the gate's leakage error channel can be effectively modeled by the following effective error channel where the imaginary terms have been removed:
\begin{equation}
\label{eq:channel}
    \Lambda_\mathrm{CZ}(\rho) = 
    \frac{1}{2} \left(\mathds{1}\rho\mathds{1} + \mathrm{CZ}\rho \mathrm{CZ} \right).
\end{equation}
Terms in the channel of the form `$\mathrm{CZ}\rho\mathds{1}$' and `$\mathds{1}\rho\mathrm{CZ}$' are annihilated when we perform the projective stabilizer measurements during quantum error correction with any stabilizer code.
One may take notice that Eq.~\ref{eq:channel} describes a Clifford error channel, not a Pauli error channel. We must be careful \textit{not} to apply any active Pauli twirling procedures to our qubits, since this would diagonalize our error channel and reduce the error correlations between our qubits. Preserving this correlation is essential for enabling delayed erasure detection when using this gate to perform stabilizer measurements in a surface code, since it ensures our leakage errors do not lead to `bad' hook errors which can reduce the effective distance of the code.  

The physical interpretation of this error channel is that leakage that occurs midway through a CZ gate can be modeled as a 50-50 probability of leakage occurring \textit{before} the CZ gate, and leakage occurring \textit{after} the CZ gate.

For completeness, we can express the total quantum channel we use to model a single CZ gate. For this, we must define the leakage operators, $\hat{L}_c=\ket{00}\bra{10}_c+\ket{00}\bra{01}_c$, the control qubit leakage jump operator and $\hat{L}_t=\ket{00}\bra{10}_t+\ket{00}\bra{01}_t$ to be the target qubit leakage jump operator. We model Pauli dephasing error gates during the gate with the channel
$\Lambda_\mathrm{dephase}(\rho)$. The total quantum channel for the gate can be written by cascading three separate error channels as
\begin{equation}
        \mathcal{E}(\rho)=\Lambda^{\text{control}}_{\text{leakage}}\circ\Lambda^{\text{target}}_{\text{leakage}}\circ\Lambda_{\text{dephase}}(\mathrm{CZ}\rho\mathrm{CZ}^\dagger),
\end{equation}
with the following channel definitions:
\begin{align}
    \Lambda^{\text{control}}_{\text{leakage}}(\rho)&= (1-p_e^c)\mathds{1}\rho\mathds{1} + p_e^c\hat{L}_c \Lambda_{\mathrm{CZ}}(\rho)\hat{L}_c^\dagger,\\
        \Lambda^{\text{target}}_{\text{leakage}}(\rho)&= (1-p_e^t)\mathds{1}\rho\mathds{1} + p_e^t\hat{L}_t \Lambda_{\mathrm{CZ}}(\rho)\hat{L}_t^\dagger,\\
        \Lambda_{\mathrm{dephase}}(\rho) &= (1-p_Z^c-p_Z^t-p_{ZZ})\mathds{1}\rho\mathds{1} \\&+ p_Z^cZ_c\rho Z_c + p_Z^tZ_t\rho Z_t \\&+ p_{ZZ} Z_tZ_c\rho Z_c Z_t.
\end{align}
For this channel definition we define $p_e^{c(t)}$ to be the probability of erasure on the control (target) qubit during a CZ gate, and $p_Z^{c(t)}$ to be the probability of a phase flip error on the control (target) qubit. Lastly, $p_{ZZ}$ is the probability of a correlated dephasing error that affects both qubits.

\subsection{Analysis of experimental results}
\label{app:leakage_propagation_analysis}

To analyze our process tomography results we apply an error analysis described in \cite{korotkov_error_2013}, which we will briefly sketch here. A quantum process can be written using the $\chi$ matrix representation as 
\begin{equation}
    \rho_{\textrm{out}} = \sum_{m,n} \chi_{mn} E_m\rho_{in} E_n^\dagger,
\end{equation}
and here we choose the Pauli basis, $\{E_m\} = \{I, X, Y, Z\}$. For our analysis, we are interested to study the \textit{deviation} from the ideal or expected process, motivating the use of the $\chi^{\textrm{(err)}}$ matrix. The total process $\chi$ can be described as the composition of a target unitary $U$ followed by an error process, $\chi^{(err)}$. Note that this error process is another valid process that can be described using the $\chi$ 
\begin{equation}
    \rho_{\textrm{out}} = \sum_{m,n} \chi^{\textrm{(err)}}_{mn} E_m U \rho_{in} U^\dagger E_n^\dagger,
\end{equation}
The error process matrix can be computed by 
\begin{equation}
    \chi^{\textrm{(err)}} = V \chi V^\dagger, V_{mn} = \mathrm{Tr}\left( E^\dagger_m E_n U\right) / d,
\end{equation}
where $d = 2^N$ with $N$ representing the number of qubits.

This error process matrix has a simple and intuitive interpretation. In the ideal case, the error process $\chi^{\textrm{(err)}}$ should equal to the identity process with a single non-zero element, $\chi_{I,I} = 1$. Any other non-zero elements indicate a deviation from the intended process. 

\section{No-jump backaction in the SWS gate}
\label{app:no_jumpbackaction}
No-jump backaction is a form of non-unitary, coherent error which arises when we detect that no excitation loss errors have occurred during our CZ gate. In essence, when we learn no loss occurred, we know it is more likely that the control qubit was in the dual-rail state $\ket{10}$ rather than $\ket{01}$, since the $\ket{10}$ state does not swap into the more lossy coupler. This amounts to a weak measurement of the control qubit in the $Z$-basis, a type of dephasing error. 

To see what limit this error places on our CZ gate fidelity, it is useful to define the quantities 
\begin{align}
    p_{\mathrm{loss}}^{a_1}&=\frac{t_{\mathrm{gate}}}{T_1^{a_1}},\\
    p_{\mathrm{loss}}^{c}&=\frac{t_{\mathrm{gate}}}{T_1^{c}}..
\end{align}

After applying the gate once, the resulting state after postselection may be approximately written as 
\begin{equation}
    \rho = \frac{\mathrm{CZ}\rho'\mathrm{CZ}}{||\rho'||},
\end{equation}
where
\begin{align}
    \rho' &= \hat{E}\rho_{\mathrm{init}}\hat{E}^\dagger,\\
    \hat{E} & = \sqrt{1-p_{\mathrm{loss}}^{a_1}}\ket{0}\bra{0}_c \otimes \mathds{1}_t + \sqrt{1-p_{\mathrm{loss}}^{c}}\ket{1}\bra{1}_c \otimes \mathds{1}_t
\end{align}
and $\hat{E}$ is the (non-trace preserving) Kraus operator describing no-jump backaction. The subscripts denote control and target qubits respectively.

The states most sensitive to this effect are ones where the control qubit is on the equator of its Bloch sphere. However, we can show that for small $|p_{\mathrm{loss}}^{a_1}-p_{\mathrm{loss}}^{c}|$, the error is approximately 
\begin{equation}
    \varepsilon_{\mathrm{NJ}} \approx \frac{(p_{\mathrm{loss}}^{a_1}-p_{\mathrm{loss}}^{c})^2}{4},
\end{equation}

which scales quadratically with the imbalance in loss rates. For our hardware parameters, we expect this effect to be small after one CZ gate, contributing with an error $<10^{-5}$ after one gate. However, after $\sim$100 consecutive gates, due to the quadratic scaling of this effect can be noticeable at the few percent level.

In any practical setting, our CZ gate will be interleaved with single qubit rotations and measurements, which have the effect of ``echoing out" the effects of no-jump backaction, ensuring that the effective error stays below $10^{-5}$.

It can be shown that a single $X$ gate on the control qubit mid-way through a repeated series of CZ gates is sufficient to completely nullify the effects of no-jump backaction. We additionally perform an $X$ gate on the target qubit as well in our quantum state tomography experiment since $X$ gates in general are useful for echoing out any sources of low frequency dephasing noise affecting our qubits. The fact that we still observe non-linear dephasing on the control qubit even with the presence of the echo pulses indicates no-jump backaction cannot be responsible for this effect. 

We now briefly describe why a single $X$ gate mid-way through our repeated CZ sequence is sufficient to nullify the effects of no-jump backaction. The intuition as to why this works is that, in the period of time that elapses from the beginning of the sequence to the final measurement, each state in the control qubit will spend the same amount of time occupying the lossy coupler mode. Thus, when we measure that photon loss has not occurred in the control qubit, we (and the environment) gain no additional information about which state the control qubit began in. 

We show this mathematically by considering an idling control dual-rail qubit, which undergoes non-unitary (and non-trace preserving) evolution due to the collapse operator 
\begin{equation}
    \hat{L} = \sqrt{\kappa}\ket{00}\bra{01}_c,
\end{equation}
where $\kappa$ models the average loss rate experienced by the $\ket{01}_c$ state as it swaps in and out of the coupler through many repeated CZ gates. 

With the Lindblad master equation we can specifically examine the no-jump part of the evolution as 
\begin{align}
    \dot{\rho} &= -\frac{1}{2}\left(\hat{L}^\dagger \hat{L}\rho + \rho\hat{L}^\dagger \hat{L} \right)\\
    &= -\frac{\kappa}{2} (\ket{01}\bra{01}_c \rho + \rho \ket{01}\bra{01}_c)\\
    & = -\frac{\kappa}{2}\rho - \frac{\kappa}{4}(Z_c \rho + \rho Z_c).
\end{align}
We can solve this equation for a ``qubitized" density matrix that describes the dual-rail codespace of the control qubit as 
\begin{equation}
    \rho(t) = \begin{pmatrix}
\rho_{00}(t) & \rho_{01}(t) \\
\rho_{10}(t) & \rho_{11}(t)
\end{pmatrix}.
\end{equation}

Solving this differential equation gives us
\begin{equation}
    \rho(t) = \begin{pmatrix}
\rho_{00}e^{-\kappa t} & \rho_{01} e^{-\kappa t/2} \\
\rho_{10}e^{-\kappa t/2} & \rho_{11}
\end{pmatrix}.
\end{equation}
This evolution describes a gradual measurement in the $Z$ basis which gradually polarizes us to the state $\ket{10}_c$, the least lossy basis state. 

Applying a single $X$ gate at time $t = \tau/2$ is sufficient to completely undo this polarization effect. After this $X$ pulse, now the state that was initially $\ket{10}$ will now swap into the coupler and the density matrix evolves under the differential equation
\begin{equation}
    \dot{\rho}
    = -\frac{\kappa}{2}\rho + \frac{\kappa}{4}(Z_c \rho + \rho z_c),
\end{equation}
which has the general solution
\begin{equation}
    \rho(t) = \begin{pmatrix}
\rho_{00} & \rho_{01} e^{-\kappa t/2} \\
\rho_{10}e^{-\kappa t/2} & \rho_{11}e^{-\kappa t}
\end{pmatrix},
\end{equation}
describing gradual polarization to the $\ket{01}_c$ state.

Thus, after evolving for time $\tau/2$ , applying the $X$ gate and then evolving for time $\tau/2$ (and finally applying another $X$ gate if desired), we are left in the state
\begin{align}
    \rho(t) = \begin{pmatrix}
\rho_{00}e^{-\kappa t} & \rho_{01} e^{-\kappa t} \\
\rho_{10}e^{-\kappa t} & \rho_{11}e^{-\kappa t}
\end{pmatrix}=e^{-\kappa t} \rho.
\end{align}
And so, the density matrix remains the same as our initial density matrix, except for multiplication by the scalar factor $e^{-\kappa t}$ which represents the probability we do not detect a loss error.

This assertion that a single $X$ gate on the control qubit mid-way through the CZ gate sequence is sufficient to completely undo the polarizing effect of no-jump back action is also supported by our QuTiP full master equation simulations.

\section{CZ induced coupler heating}
\label{sec:coupler_heating}
\begin{figure}[ht]
\centering
\includegraphics[width=3 in]{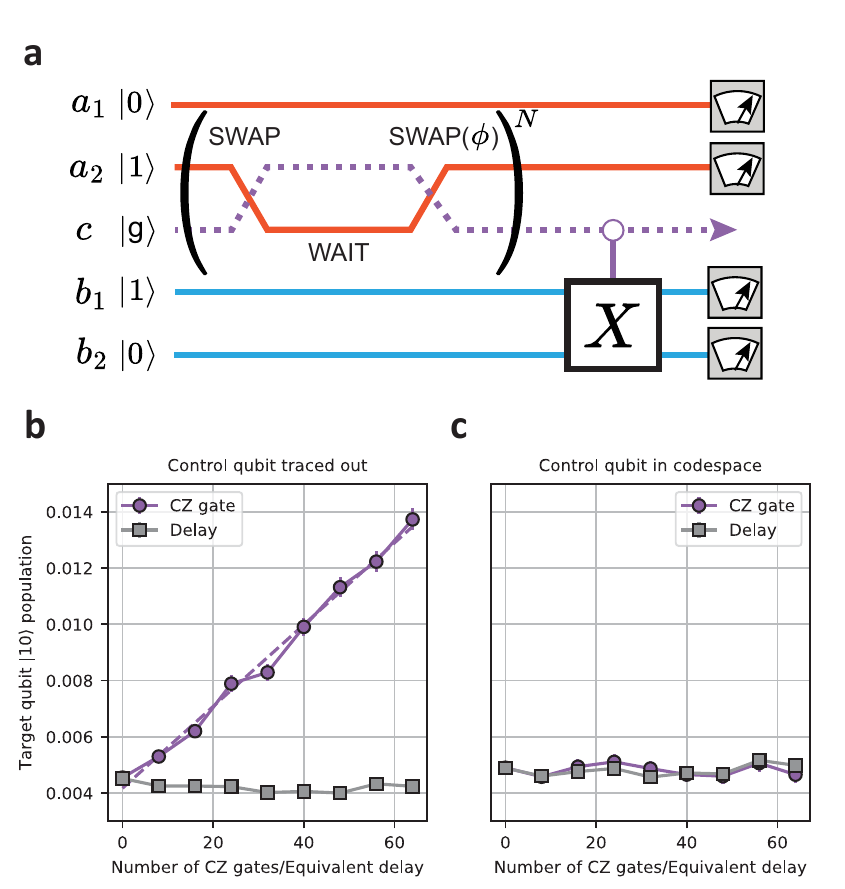}
\caption{
    \label{fig_coupler_heating}
    \textbf{Residual coupler excitation}
    \textbf{a.} Pulse level protocol used to map the residual coupler population into target qubit.
    \textbf {b.} Target qubit $|01 \rangle$ population as a function of number of CZ gates in the sequence or delay when the control qubit measurement outcome are traced out  
    \textbf{c.} Same as \textbf{b} but when control qubit measurement outcomes are postselected to be in the codespace.
    \label{coupler_heating}
}
\end{figure}

Since the transmon coupler mode $c$ is temporarily populated during the application of the CZ gate it is important to ensure that all the population is transferred back to the control dual-rail before the gate ends, failing to do so will dephase the target qubit. In this section we describe the experimental protocol to measure the residual coupler excitation at the end of the CZ gate. We do this by preparing our system in the state $|01g10 \rangle$ and repeatedly applying the CZ gate followed by a slow selective $X$ gate on the target qubit. This selective $X$ gate only swaps excitation between cavities $b_1$ and $b_2$ when the coupler is in the $|g \rangle$ state and implements an identity on the target qubit when the coupler is in the $|e \rangle$ state. This pulse sequence maps the residual coupler excitation onto the target qubit $| 10 \rangle$ state population. At the end of the sequence we measure the two dual-rail qubits. We process the data in two ways. First, we trace out the control qubit and measure the population of the target qubit in the state $|10 \rangle$. This encodes the unpostselected residual coupler excitation at the end of $N$ CZ gate applications. We observe a small but measurable increase in the coupler population of  ~$0.014 \% $ per gate application (see Fig \ref{coupler_heating}  b). Next, we use the same dataset but postselect on the measurement outcomes such that the control qubit is assigned to be in the codespace. In this case we observe no measurable increase of the target qubit $|10 \rangle$  population after 64 applications of the gate. Its important to note here that because of the excitation preserving property of the gate whenever we inadvertently leave the coupler in the excited state we will necessarily observe the control qubit in the vacuum sate. This allows us to postselect on this error channel. Lastly, we also compare this measurement to the case when we replace the CZ gate with a delay of equivalent duration and find that we see no measurable difference between the two sequences when the control qubit is postselected to be in the code space. 

\section{Correlated leakage and coupler excitation in the SWS gate}
Although rare, there is a mechanism by which photon loss on a target qubit midway through an SWS gate can result in both the control and target qubit ending in the vacuum state, and an excitation remaining in the coupler. We refer to this as a correlated error because a single loss mechanism can result in a double erasure. This effect is most pertinent when $g_{ac}\sim\chi_{bc}$, when the control qubit does not fully swap its photon from $a_2$ into the coupler mode. Photon loss from $b_1$ at unknown time dephases any excitations in the coupler, and in this parameter regime, there is also a chance the coupler excitation is not fully swapped back into the cavity mode $a_2$, resulting in the `double-erasure' state $\ket{00e00}$. 

For our hardware parameters, we expect to end up in the double-erasure state which  occurs with probability $1.3\times10^{-5}$ per CZ gate, according to our QuTiP simulations. Equivalently, if photon loss occurs in the target qubit during the gate, the excitation is expected to remain stuck in the coupler with probability $1.4\%$. This probability scales quadratically with the ratio $|\chi_{bc}|/g_{ac}$ and is not expected to limit gate performance.  

A related error, dephasing in the coupler mode midway through the SWS, can also put us in the detectable leakage states $\ket{00e10}$ or $\ket{00e01}$. This arises because dephasing of the coupler mode no longer guarantees that second SWAP has the correct phase. Like the aforementioned double erasure, this error is also suppressed quadratically with the ratio $|\chi_{bc}|/g_{ac}$, which serves as another reason to increase $g_{ac}$ in future implementations. 

With our QuTiP master equations, we predict the system leaks to these two states with probability $6.5\times10^{-5}$ per gate, assuming a coupler dephasing time of $1001 \upmu$s. The total probability the coupler is predicted to end in the excited state is $8.0\times10^{-5}$ per gate, comparable to what we measure in experiment.

\section{Measuring bit-flip errors}

\begin{figure}[ht]
\centering
\includegraphics{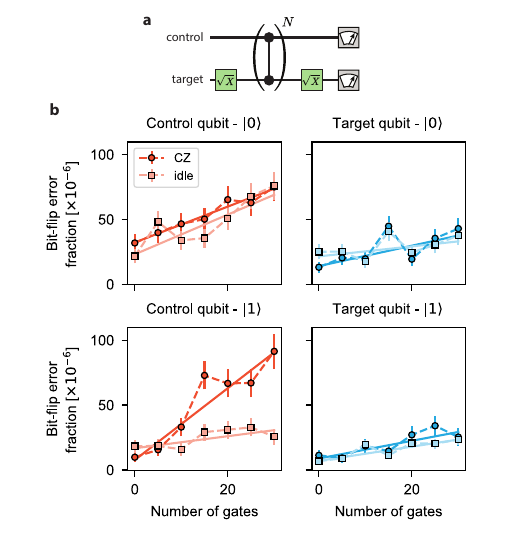}
\caption{
    \label{fig_bit_flip}
    \textbf{Measuring bit-flip errors during CZ gate}
    \textbf {a.} Circuit used to measure bit-flip errors 
    \textbf{b.} Results of bit-flip measurements for the various combinations of initial logical states of the target and control dual-rails, compared to the idling states when no CZ gates are performed. The extracted linear trends (solid lines) are used to produce the quantities presented in Fig.~\ref{fig3}f.
}
\end{figure}

Figure~\ref{fig_bit_flip} presents additional data, obtained with other variations of the initial logical states of the target and control qubits, with the same measurement protocol described in the main text. The fits to the linear slopes (solid lines) are used to extract the quantitites plotted in Fig.~\ref{fig3}f. With the exception of the state where an excitation is swapped into the coupler, the rates are indistinguishable from the idling bit flip rates. As in \cite{chou_superconducting_2024}, the quantities we extract are likely only an apparent rate, due to the finite leakage discrimination fidelity of the end-of-the-line measurements, multiplied by the increasing number of erasures with time.

\section{Derivation of gate parameters}
In our gate, there are three parameters that must be calibrated. These are the duration of each of the cavity-coupler `SWAP' pulses, $t_{\mathrm{SWAP}}$, the delay time between these two pulses, $t_{\mathrm{wait}}$, and the phase of the second SWAP pulse, relative to the first, which we denote as $\phi$.

In this section we derive the theoretical values for these three parameters required to implement a unitary that is locally equivalent to a $\mathrm{CZ}$ gate on the dual-rail subspace, up to local $Z$ rotations. We find these parameters for the assumption that we have square, instantaneous pulses and that our pulse Hamiltonian takes the form
\begin{equation}
    \mathcal{H}/\hbar = \frac{1}{2}\left(g_{ac}(t) a_2^\dagger c + g_{ac}^*(t)a_2 c^\dagger\right) + \chi_{bc} b_1^\dagger b_1 c^\dagger c
\end{equation}
In the actual experiment, these values need to be calibrated and will differ from the theoretical values we are about to derive due to finite pulse ramp times and extra Hamiltonian terms such as cavity-cavity cross Kerr. 

Nevertheless, deriving these parameters is useful for numerical simulation, and for understanding the physics of the gate. 

One thing to note is that for a given set of hardware parameters, $(g_{ac},\chi_{bc})$, there are multiple sets of pulse parameters, $(t_{\mathrm{SWAP}},t_{\mathrm{wait}}, \phi)$ which can enact a $\mathrm{CZ}$ gate. To simplify this, we can choose to fix one of the parameters. In \cite{RosenblumGao2018}, Rosenblum {\it et al.} choose to fix $\phi=\pi$, and calibrate the two pulse timing parameters. We instead choose to fix $t_{\mathrm{SWAP}}=\pi/g_{ac}$, taking $g_{ac}$ to be real and positive. (Experimentally, this means we only need to calibrate a delay time and a phase, the easiest parameters to sweep). 

Given this constraint, the analytical expressions for $t_{\mathrm{wait}}$ and $\phi$ are
\begin{align}
\label{eq:solutions}
    t_{\mathrm{wait}}&=\frac{\pi}{|\chi_{bc}|} - \frac{\pi}{g_{ac}},\\
    \phi&= \chi_{bc} t_{\mathrm{wait}} \\
    &\quad + 2\arctan{\left(-\frac{\sqrt{g_{ac}^2+\chi_{bc}^2}}{\chi_{bc}}\cot{\left(\frac{\pi\sqrt{g_{ac}^2+\chi_{bc}^2}}{2g_{ac}}\right)}\right)},
\end{align}
derived below.
With these parameters, the gate can be modeled through three different piece-wise Hamiltonians:
\begin{align}
\label{eq:piecewise_H}
    \mathcal{H}_1/\hbar &= \frac{g_{ac}}{2} (a_2^\dagger c + a_2 c^\dagger) + \chi_{bc} b_1^\dagger b_1 c^\dagger c \\
    \mathcal{H}_2/\hbar &= \chi_{bc} b_1^\dagger b_1 c^\dagger c\\
    \mathcal{H}_3/\hbar &=\frac{g_{ac}}{2} (e^{i\phi}a_2^\dagger c + e^{-i\phi}a_2 c^\dagger) + \chi_{bc} b_1^\dagger b_1 c^\dagger c
\end{align}

acting for times, $t_{\mathrm{SWAP}}$, $t_{\mathrm{wait}}$, $t_{\mathrm{SWAP}}$ respectively

It also follows that the total gate duration is given by
\begin{equation}
    t_{\mathrm{gate}} = t_{\mathrm{wait}} + 2t_{\mathrm{SWAP}} = \frac{\pi}{|\chi_{bc}|} + \frac{\pi}{g_{ac}}
\end{equation}

The unitary on the dual-rail space that is implemented is 
\begin{equation}
    U_\mathrm{gate} = Z_1(\pi-\phi)Z_2(\pi)U_{\mathrm{CZ}}
\end{equation}
where we use the usual convention
\begin{equation}
    U_{\mathrm{CZ}} = \begin{pmatrix}
1 & 0 & 0 & 0 \\
0 & 1 & 0 & 0\\
0 & 0 & 1 & 0 \\
0 & 0 & 0 & -1 
\end{pmatrix}
\end{equation}

\subsection{Deriving $t_{\mathrm{wait}}$}
During the gate, only the dual-rail state $\ket{0110}$ gains an entangling phase, via the dispersive interaction. To find the wait time, we must tally up the entangling phase accumulated during each of the three piece-wise Hamiltonians. The total entangling phase (which must be $\pi$ for a $\mathrm{CZ}$ gate) is then given by 
\begin{equation}
\label{eq:tot_ent_phase}
    \phi_{\mathrm{ent}} = 2\phi_{\mathrm{SWAP}} + \phi_{\mathrm{wait}} = \pi
\end{equation}
where $\phi_{\mathrm{wait}}$, the phase accumulation during the delay time is simply $\phi_{\mathrm{wait}} = \chi t_{\mathrm{wait}} $. Finding the entangling phase contribution from $\mathcal{H}_1$ and $\mathcal{H}_3$ is a little trickier, but can be found by considering a `qubitized' Hamiltonian of the form

\begin{equation}
    \mathcal{H}_q = \frac{g_{ac}}{2}\begin{pmatrix}
0 & 1 \\
1 & 0 
\end{pmatrix}+\chi_{bc} \begin{pmatrix}
0 & 0 \\
0 & 1 
\end{pmatrix} 
\end{equation}
which describes the dynamics of the dual-rail state $\ket{01g10}$ swapping to the state $\ket{00e10}$ in a two-level manifold. 
If we were to enact this Hamiltonian for time $\tau=2\pi/\Omega$, where $\Omega=\sqrt{g_{ac}^2+\chi_{bc}^2}$ is the detuned Rabi rate, then we would find the unitary this enacts is
\begin{equation}
e^{i\mathcal{H}_q\tau} = e^{i(\pi-\chi_{bc}\tau/2)}\mathds{1}
\end{equation}
Usually this phase is ignored as a `global phase', but in our case, because this is a phase on just one of the dual-rail basis states, it is not global, and it contributes to a measurable phase on our dual-rail qubit. The phase contribution due to $\chi$ from a `complete orbit' of a detuned Rabi oscillation is $\chi_{bc}\tau/2$. (The $\pi$ offset does not contribute to the entangling phase). If we instead were to enact $\mathcal{H}_q$ for time $t=t_{\mathrm{SWAP}} = \pi/g_{ac}$, we would instead acquire a smaller phase, scaled by the fraction of the `orbit' we complete:
\begin{equation}
\label{eq:ent_swap_phase}
   \phi_{\mathrm{SWAP}} =\frac{\chi\tau}{2} \frac{t_{\mathrm{SWAP}}}{\tau} = \frac{\chi t_{\mathrm{SWAP}}}{2}
\end{equation}
This is equivalent to the entangling phase we acquire from enacting $\mathcal{H}_1$ for time $t_{\mathrm{SWAP}}$. The contribution to the total entangling phase from $\mathcal{H}_3$ is also identical.

From \ref{eq:ent_swap_phase} and \ref{eq:tot_ent_phase} we can write
\begin{equation}
    2 \frac{\chi_{bc} t_{\mathrm{SWAP}}}{2} + \chi t_{\mathrm{wait}} = \pi
\end{equation}
and is easily rearranged to give
\begin{equation}
    t_{\mathrm{wait}}=\frac{\pi}{|\chi_{bc}|} - \frac{\pi}{g_{ac}}
    \label{twait}
\end{equation}
which is the first expression we wanted to derive and takes a surprisingly simple form.

\subsection{Deriving $\phi_{\mathrm{SWAP}}$}

\begin{figure}[ht]
    \centering
    \includegraphics[width=0.6\linewidth]{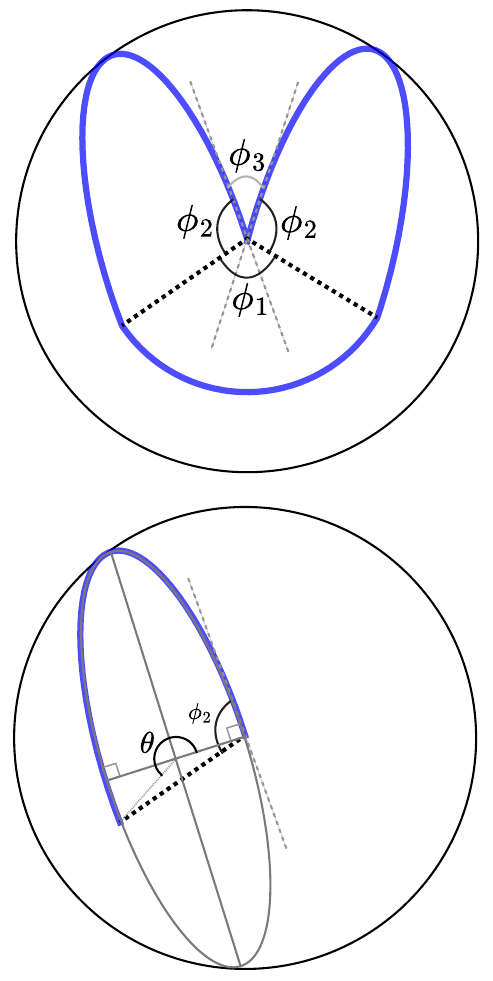}
    \caption{(Top) Top-down view of the Bloch sphere trajectory of the dual-rail basis state $\ket{0_L1_L}$, projected into the XY plane. The Bloch sphere is defined as in the left panel of Fig.~\ref{fig:bloch_traj}. Finding the analytical expression for $\phi_{\mathrm{SWAP}}$ amounts to calculating $\phi_3$ on this diagram. (Bottom) Top down view of the same trajectory, focussing on the evolution during $\mathcal{H}_1$, with useful angles marked to find $\phi_2$.}
    \label{fig:swap_back_phase}
\end{figure}

As may be expected from the form of the analytical expression for $\phi_{\mathrm{SWAP}}$ in Eq. \ref{eq:solutions}, this derivation will require some geometry and trigonometry. It is helpful to refer to Fig.~\ref{fig:swap_back_phase}. The angle $\phi_3$ is related to $\phi_{\mathrm{SWAP}}$ as $\phi_3 = \pi - \phi_{\mathrm{SWAP}}$. We also know the value of $t_{\mathrm{wait}}$ and so $\phi_2 = \chi_{bc}t_{\mathrm{wait}}$. If we can calculate $\phi_2$ then we can use the fact $\phi_1 + 2\phi_2 + \phi_3 = 2\pi$ to calculate $\phi_3$ and thus $\phi_{\mathrm{SWAP}}$. We start by noting that the evolution during $\mathcal{H}_1$ (and $\mathcal{H}_3$) is an ellipse in the XY plane. On the Bloch sphere, these amount to detuned Rabi oscillations but when projected onto the XY plane these become elliptical. We define the major axis of this ellipse to be $2y_0$ and the minor axis to be $2x_0$. It can be shown that the ratio $y_0/x_0=\chi_{bc}/\Omega$.

If we were to center this ellipse at the origin (see Fig.~\ref{fig:swap_back_phase}), then the $(x,y)$ coordinates can be parameterized by angle $\theta(t)$, where a complete orbit of the detuned Rabi oscillation is completed at $t = 2\pi/\Omega$. The $(x,y)$ coordinates are then given by
\begin{align}
    x(t) &= x_0 \cos(\theta(t))\\
    y(t) &= y_0 \sin(\theta(t)).
\end{align}

However, since we only activate $\mathcal{H}_1$ for time $t=t_{\mathrm{SWAP}}$, our partially complete orbit ends at $\theta(t) = \Omega t_{\mathrm{SWAP}} = \theta'$. We can now simply write down our end point, but will also shift the origin of our coordinate system such that $(0,0)$ coincides with the beginning of the trajectory (i.e. the origin is the center of the XY plane of our Bloch sphere projection). Now, the endpoints of our trajectory are written as 
\begin{align}
    x' &= x_0 (1-\cos(\Omega t_{\mathrm{wait}}))\\
    y' & = y_0\sin(\Omega t_{\mathrm{wait}}). 
\end{align}
We can use this to find 
\begin{equation}
    \phi_2 = \frac{\pi}{2}+\tan^{-1}\left(\frac{|y'|}{|x'|}\right),
\end{equation}
and so we may finally arrive at
\begin{align}
    \phi_{\mathrm{SWAP}} &= \phi_1 + 2\phi_2 - \pi\\
    &=\chi_{bc}t_{\mathrm{wait}}+2\tan^{-1}\left(\frac{|y'|}{|x'|}\right)\\
    &=\chi_{bc}t_{\mathrm{wait}}+2\tan^{-1}\left(\frac{-y_0\sin(\Omega t_{\mathrm{SWAP}})}{x_0(1-\cos(\Omega t_{\mathrm{SWAP}}))}\right)\\
    &=\chi_{bc}t_{\mathrm{wait}}+2\tan^{-1}\left(-\frac{\Omega}{\chi}\cot\left(\frac{\Omega t_{\mathrm{SWAP}}}{2}\right)\right),
\end{align}
where we have used the condition $\Omega t_{\mathrm{SWAP}} > \pi$.

\section{QuTiP Master equation simulations and CZ gate error budget}

\begin{figure}[ht]
    \centering
    \includegraphics[width=0.9\linewidth]{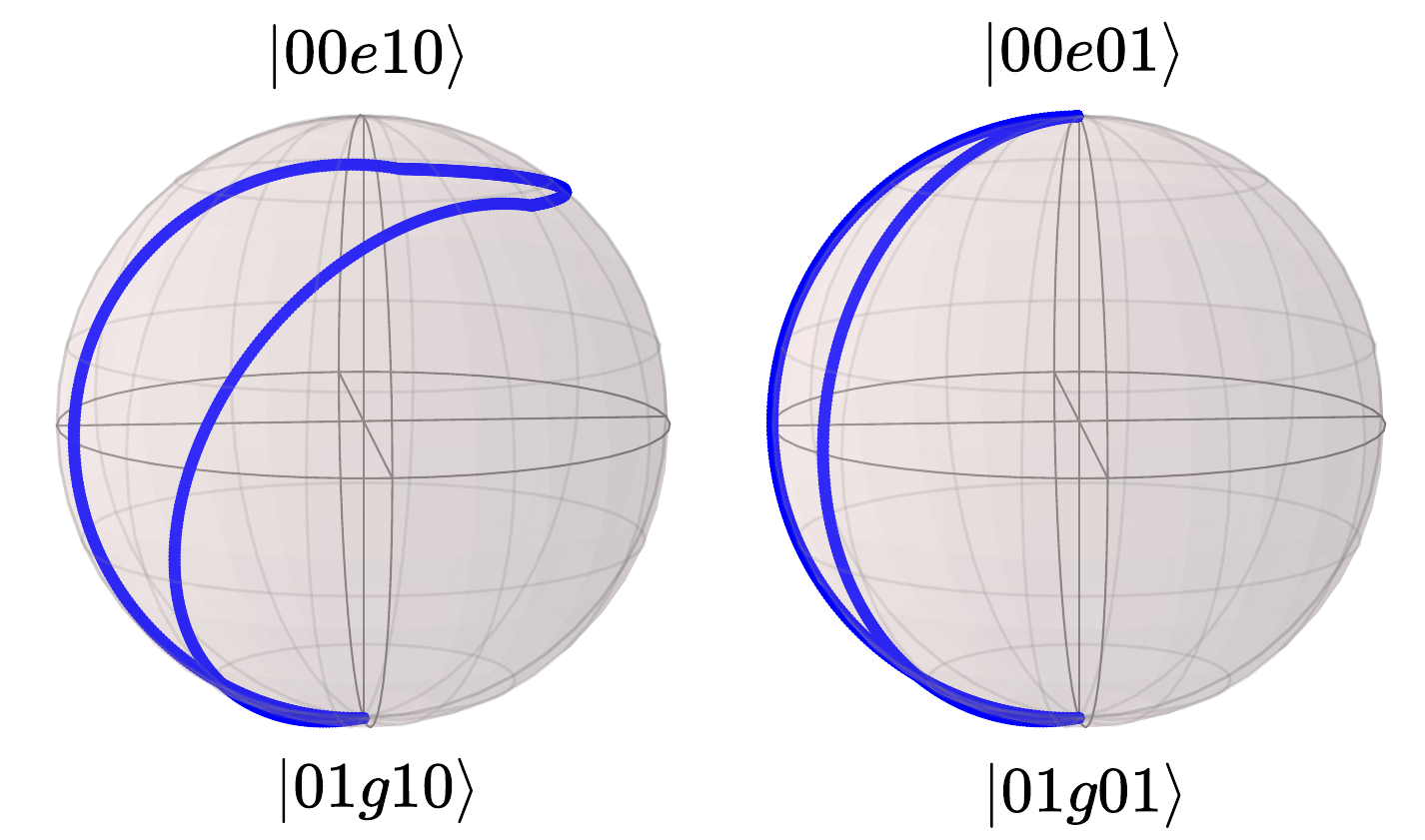}
    \caption{Evolution of the dual-rail basis states $\ket{0_L1_L}$ (left) and $\ket{0_L0_L}$ (right) during the SWS gate, obtained from our QuTiP  simulations. }
    \label{fig:bloch_traj}
\end{figure}

\begin{table*}[ht!]
\label{tab:sim_error_budget}
\begin{tabular}{c|l|c|c}
Error channel & Error state(s) after gate & Probability (QuTiP) & Probability (at short gate depth) \\
\hline
\hline
Photon loss on control DR & $\ket{00g01},\ket{00g10}$ & 0.337\% & 0.400(4)\%  \\
Photon loss on target DR & $\ket{10g00},\ket{01g00}$ & 0.0982\% & 0.096(4)\% \\
Photon loss on both DRs & $\ket{00g00}$ & 0.0003\% & -  \\

Control DR excitation stuck in coupler mode & $\ket{00e10},\ket{00e01}$ & 0.006\% & - \\
Only one excitation remains in coupler mode  & $\ket{00e00}$ & 0.0014\% & - \\
Control DR dephasing & $Z_c\ket{\psi_{\mathrm{ideal}}}$ & 0.0169\% &  0.039(1)\%\\
Target DR dephasing & $Z_t\ket{\psi_{\mathrm{ideal}}}$ & 0.0048\% & 0.0112(9)\% \\
Control and target DR dephasing & $Z_cZ_t\ket{\psi_{\mathrm{ideal}}}$ & 0.0001\% & - \\
No error & $\ket{\psi_{\mathrm{ideal}}}$ & 99.5351\% &- 
\end{tabular}
\caption{Estimated error budget for a single CZ gate, based on QuTiP simulations and measured hardware coherence times. The rightmost column compares to experimentally measured values at short numbers of gate repeates.}
\end{table*}

We simulate expected gate performance by simulating evolution under the three piecewise Hamiltonians described in \ref{eq:piecewise_H} (i.e. with the square pulse simplification), modelling all 5 modes $a_1, a_2, c, b_1, b_2$ and their respective collapse operators, modelling white-noise dephasing and excitation loss. The loss rates for these collapse operators are obtained from our measured hardware coherence times. 

These simulations capture the effects of no-jump back-action, expected Z-frame updates, correlated errors, and verify that our expressions for derived gate parameters in \ref{eq:solutions} are exact. 

We can use our QuTiP simulations in the absence of decoherence to examine the intermediate evolution of dual-rail basis states during the SWS gate, as shown in Fig.~\ref{fig:bloch_traj}. The $\ket{0_L}\ket{1_L}$ completes a closed trajectory within the $\ket{01g10}-\ket{00e10}$ manifold, as shown on the Bloch sphere, gaining a  total phase shift of $\pi +\phi_{\mathrm{SWAP}}$, and population returns exactly to the initial $\ket{01g10}$ state. In contrast, the evolution of the $\ket{0_L1_L}$ dual-rail state is much simpler, whereby the state $\ket{01g01}$ swaps completely into the coupler, and then is completely swapped back to gain a geometric phase $\phi_\mathrm{SWAP}$. The net result is an entangling phase of $\pi$ and a determinstic Z-phase of $\phi_\mathrm{SWAP}$ on the control qubit, a unitary that is equivalent to a CZ gate. The evolution of the dual-rail states $\ket{1_L0_L}$ and $\ket{1_L1_L}$ is trivially identity since there are no photons in mode $a_2$ to swap into the coupler, and thus no phase accumulation.

We can use this QuTiP simulation to numerically simulate the expected error budget for our gate, given measured hardware coherence times, and compare it against some of our measured gate errors. For this, we use the coherence times listed in Tab.~\ref{tab:system_properties} and \ref{tab:DR_coherences}, taking the longer echo times as our dephasing times. The simulated error budget is shown in Tab.~\ref{tab:sim_error_budget} and allows us to estimate the probabilities for different CZ gate error channels that are too small to measure directly in experiment.

\section{Fundamental limits to SWS gate performance}

\begin{table}[ht]
    \centering
\begin{tabular}{c|c}
    SWS error rate & Error scaling \\
    \hline
    $p_e^\mathrm{control}$ & $\frac{1}{G}\frac{1}{\alpha_c T_c^1}$ \\
    $p_e^\mathrm{target}$ & $\frac{1}{\alpha_c T_c^1}$ \\ 
    $p_Z^\mathrm{control}$ & $\frac{1}{G}\frac{1}{\alpha_c T_c^\phi}$ \\
    $p_Z^\mathrm{target}$ & $G\times \frac{1}{\alpha_c T_c^\phi}$ \\  
\end{tabular}
    \caption{Fundamental limits to the SWS if we assume inherited coupler decoherence is responsible for all cavity decoherence. $G$ typically takes a value between 0.01-0.05 in our hardware.}
    \label{tab:fund_scaling}
\end{table}

The SWS gate uses the dispersive interaction of a transmon coupler as the source of entanglement. This has many similarities to schemes where dispersively coupled transmon ancillae are used for quantum control of cavity modes. As such, one may suspect that transmon decoherence would set a fundamental limit on our gate fidelity. 

To investigate this limit, we consider a hypothetical scenario where only the transmon coupler suffers from loss and dephasing noise, whilst all cavity modes have no intrinsic noise (infinite cavity $T_1$ and $T_\phi$). In this limit, the cavities dispersively coupled will inherit both dephasing noise and loss from the coupler via the inverse Purcell effect. This can be quantified by defining the hybridization factors

\begin{align}
G_a &= \left(\frac{g^0_{ac}}{\Delta_{ac}}\right)^2 \\ 
G_b &= \left(\frac{g^0_{bc}}{\Delta_{bc}}\right)^2, 
\end{align}
where $g^0_{ac}$ is the vacuum Rabi coupling and $\Delta_{ac}$ is the mode detuning between modes $a_2$ and $c$ (in the context of a Jaynes-Cummings Hamiltonian) and similarly for $g^0_{bc}$ $\Delta_{ac}$ for modes $b_1$ and $c$. $G_a$ and $G_b$ also represent the energy participation of the dressed cavity modes in the coupler mode. 

Under an assumed white noise dephasing model, we expect the following decoherence times inherited by the cavity modes from the coupler:
\begin{align}
    T^1_{a_2} &= \frac{T^1_c}{G_a}\\
    T^1_{b_1} &= \frac{T^1_c}{G_b}\\
    T^\phi_{a_2} &= \frac{T^\phi_c}{(G_a)^2}\\
    T^\phi_{b_1} &= \frac{T^\phi_c}{(G_b)^2}
\end{align}
The goal is to calculate the SWS gate errors $p_e^\mathrm{control}, p_e^\mathrm{target}, p_z^\mathrm{control}, p_z^\mathrm{target}$ based on these coherence times and also the duration of the SWS gate, which is limited to $\tau = \pi/|\chi_{bc}|$. 

It will be insightful to rewrite this $\chi$ in relation to the anharmonicity of the transmon coupler, which is approximately $\chi_{bc} \approx 2 G_b \alpha_c$ where $\alpha_c$ is the anharmonicity of the coupler.

Now, we simply use the ratio of gate time to coherence time to calculate the SWS error limits. We omit prefactors of order unity and assume $G\sim G_a \sim G_b$. For our hardware, $G$ typically varies between 0.01 and 0.05. As an example,
\begin{equation}
p_e^\mathrm{target} \sim \frac{\tau}{T_{b_1}^1}\approx \frac{\pi}{|\chi_{bc}| T_{b_1}^1}\approx\frac{G_b\pi}{2G_b\alpha_c T_c^1} \sim \frac{1}{\alpha_c T_1^c}.   
\end{equation}
For the control qubit, we must take into account the fact that the excitation is swapped into the coupler for most of the gate, and so experience coupler decoherence more directly. As an example,
\begin{equation}
    p_Z^{\mathrm{control}}\sim \frac{\tau}{T_c^\phi}\approx\frac{\pi}{|\chi_{bc}| T_c^\phi}\sim \frac{1}{G}\frac{1}{\alpha_cT^{\phi}_c}
\end{equation}
The results of this analysis for the other three error channels is shown in Table \ref{tab:fund_scaling}.

Factors of the form $1/\alpha_c T^\phi_c$ are indicative of transmon gate fidelities at the decoherence limit (i.e. typical single qubit gate fidelities on transmons). The fact that $p_Z^\mathrm{target}$ may be several orders of magnitude better than the transmon decoherence limit makes the SWS gate extremely promising for flag-qubit or stabilizer measurement circuits, with the maximum bias between $p_Z^\mathrm{target}$ to $p_Z^\mathrm{control}$ set by $1/G^2$. We should expect $p_Z^\mathrm{target}$ to further improve as we reduce extrinsic sources of cavity dephasing (e.g. dephasing inherited from the ancilla transmons).

\end{document}